# Adaptive Bayesian Radio Tomography

Donghoon Lee, *Student Member, IEEE,* Dimitris Berberidis, *Student Member, IEEE,*
and Georgios B. Giannakis, *Fellow, IEEE*

*Abstract*—Radio tomographic imaging (RTI) is an emerging technology to locate physical objects in a geographical area covered by wireless networks. From the attenuation measurements collected at spatially distributed sensors, radio tomography capitalizes on spatial loss fields (SLFs) measuring the absorption of radio frequency waves at each location along the propagation path. These SLFs can be utilized for interference management in wireless communication networks, environmental monitoring, and survivor localization after natural disaster such as earthquakes. Key to success of RTI is to model accurately the shadowing effects as the bi-dimensional integral of the SLF scaled by a weight function, which is estimated using regularized regression. However, the existing approaches are less effective when the propagation environment is heterogeneous. To cope with this, the present work introduces a piecewise homogeneous SLF governed by a hidden Markov random field (MRF) model. Efficient and tractable SLF estimators are developed by leveraging Markov chain Monte Carlo (MCMC) techniques. Furthermore, an uncertainty sampling method is developed to adaptively collect informative measurements in estimating the SLF. Numerical tests using synthetic and real datasets demonstrate capabilities of the proposed algorithm for radio tomography and channel-gain estimation.

## I. INTRODUCTION

Tomographic imaging is a technique widely appreciated by natural sciences, notably in medical imaging [29]. The principles underpinning radio tomographic methods have been carried over to construct underlying *spatial loss fields* (SLFs), which are maps quantifying the attenuation experienced by electromagnetic waves in radio frequency (RF) bands at every spatial position [24]. To this end, pairs of collaborating sensors are deployed over the area of interest to estimate the attenuation introduced by the channel between those pairs of sensors. Different from conventional methods, radio tomography relies on *incoherent* measurements containing no phase information; see also [17] for another application of incoherent measurements to cognitive radio networks. Such simplification saves costs incurred for synchronization that is necessary to calibrate phase differences among waveforms received at different sensors.

Parts of this work will be presented at the IEEE International Conference on Acoustics, Speech and Signal Processing, held in Calgary, Alberta, Canada, during Apr. 15-20, 2018.
D. Lee, D. Berberidis, and G. B. Giannakis are with the Department of Electrical and Computer Engineering and the Digital Technology Center, University of Minnesota, Minneapolis, MN 55455, USA. Emails: {leex6962, bermp001, georgios}@umn.edu.
The work in this paper was supported by grants NSF grants 1442686, 1508993, 1509040.

SLFs are instrumental in various problems including radio tomography [32] and channel-gain cartography [16]. The absorption captured by the SLF allows one to discern objects located in the area of interest, thus enabling radio tomographic imaging (RTI). Benefiting from the ability of RF waves to penetrate physical structures such as trees or buildings, RTI provides a means of device-free passive localization [33], [34], and has found diverse applications in disaster response situations for e.g., detecting individuals trapped in buildings or smoke [31]. SLFs are also useful in channel-gain cartography to provide channel-state information (CSI) for links between arbitrary locations even where no sensors are present [16]. Such maps can be employed in cognitive radio setups to control the interference that a secondary network inflicts to primary users that do not transmit–a setup encountered with television broadcast systems [36], [5], [15]. The non-collaborative nature of these primary users precludes training-based channel estimation between secondary transmitters and primary receivers. Other applications of channel-gain maps include network planning, and interference management in cellular networks.

The key premise behind RTI is that spatially close radio links exhibit similar shadowing due to the presence of common obstructions. This shadowing correlation is related to the geometry of objects present in the area waves propagate through [24], [1]. As a result, shadowing is modeled as the weighted line integral of the underlying two-dimensional SLF. The weights in the integral are determined by a function depending on the transmitter-receiver locations [24], [9], [27], which models the SLF influence on the shadowing over a link between those transceivers. Inspired by this SLF model, various tomographic imaging methods were proposed [32], [31], [30], [14]. To detect locations of changes in the propagation environment, the difference between the SLF at consecutive time slots was employed [32], [30]. To cope with multipath fading in a cluttered environment, multiple channel measurements were utilized to enhance localization accuracy [14]. Although these are calibration-free approaches, they cannot reveal static objects in the area of interest. It is also possible to replace the SLF with a label field indicating presence (or absence) of objects in motion on each voxel [31], and leverage the influence moving objects on the propagation path have on variance in RSS measurements. On the other hand, the SLF was directly reconstructed in [9] to depict the static structure in the area of interest, but calibration was necessary by using measurements collected without the structure.

A different body of works inspired by the SLF model is available for channel-gain cartography [16], [27], [4], [18]. Linear interpolation techniques such as kriging were employed to estimate shadowing effects based on spatially correlated measurements [4], and the spatio-temporal dynamics were tracked by using Kalman filtering approaches [16]. SLFs with 'regular patterns' of objects have also been modeled as a superposition of a low-rank matrix plus a sparse matrix capturing structure irregularities [18]. While the aforementioned methods rely on heuristic criteria to choose the weight function, [27] provides a suite of blind algorithms to learn the weight function using a non-parametric kernel regression method.

Conventionally, the SLF is learned via regularized least-squares (LS) methods tailored to the propagation environment [9], [18], [30]. However, these approaches are less effective when the propagation environment exhibits heterogeneous characteristics. To account for environmental heterogeneity, the novel method here leverages the Bayesian framework to learn the piecewise homogeneous SLF through a hidden Markov random field (MRF) model [12], which captures spatial correlations of neighboring regions exhibiting related statistical behavior. Efficient field estimators will be derived by using Markov chain Monte Carlo (MCMC) sampling [8], which is a powerful tool for Bayesian inference when analytical solutions of the minimum mean-square error (MMSE) or the maximum a posteriori (MAP) estimators are not available. Furthermore, hyperparameters are estimated as well, instead of being fixed a priori.

Besides accounting for heterogeneous propagation, another contribution here is an adaptive data acquisition technique, with the goal of reducing SLF uncertainty, by cross-fertilizing ideas from the fields of experimental design [6] and active learning [20]. The conditional entropy of the SLF is considered as an uncertainty measure in this work, giving rise to a novel data acquisition criterion. Although such criterion is intractable especially when the size of the SLF is large, its efficient proxy can be obtained thanks to the availability of posterior samples from the proposed MCMC-based algorithm.

The rest of the paper is organized as follows. Sec. II reviews the radio tomography model and states the problem. The Bayesian model and the resultant field reconstruction are the subjects of Sec. III. Numerical tests with synthetic as well as real measurements are provided in Sec. IV. Finally, Sec. V summarizes the main conclusions.

*Notations:* Bold uppercase (lowercase) letters denote matrices (column vectors). Calligraphic letters are used for sets; $\mathbf{I}_n$ is the $n \times n$ identity matrix; while $\mathbf{0}_n$ and $\mathbf{1}_n$ denote $n \times 1$ vectors of all zeros and ones, respectively. Operators $(\cdot)^\top$ and $\mathrm{tr}(\cdot)$ represent the transpose and trace of a matrix $\mathbf{X} \in \mathbb{R}^{N_x \times N_y}$, respectively; $|\cdot|$ is used for the cardinality of a set, and the magnitude of a scalar; and $\mathrm{vec}(\mathbf{X})$ produces a column vector $\mathbf{x} \in \mathbb{R}^{N_x N_y}$ by stacking the columns of a matrix one below the other ($\mathrm{unvec}(\mathbf{x})$ denotes the reverse process). For a vector $\mathbf{y} \in \mathbb{R}^n$ and an $n \times n$ weight matrix $\mathbf{\Delta}$, the weighted norm of $\mathbf{y}$ is $\|\mathbf{y}\|_\mathbf{\Delta}^2 := \mathbf{y}^\top \mathbf{\Delta} \mathbf{y}$.

## II. BACKGROUND AND PROBLEM STATEMENT

Consider a set of sensors deployed over a two-dimensional geographical area indexed by a set $\mathcal{A} \subset \mathbb{R}^2$. After averaging out small-scale fading effects, the channel-gain measurement over a link between a transmitter located at $\mathbf{x} \in \mathcal{A}$ and a receiver located at $\mathbf{x}' \in \mathcal{A}$ can be represented (in dB) as

$$g(\mathbf{x}, \mathbf{x}') = g_0 - \gamma 10 \log_{10} d(\mathbf{x}, \mathbf{x}') - s(\mathbf{x}, \mathbf{x}') \quad (1)$$

where $g_0$ is the path gain at unit distance; $d(\mathbf{x}, \mathbf{x}') := \|\mathbf{x} - \mathbf{x}'\|$ is the Euclidean distance between the transceivers at $\mathbf{x}$ and $\mathbf{x}'$; $\gamma$ is the pathloss exponent; and $s(\mathbf{x}, \mathbf{x}')$ is the attenuation due to shadow fading. For radio tomography, a tomographic model for the shadow fading is [24], [9], [18]

$$s(\mathbf{x}, \mathbf{x}') = \int_\mathcal{A} w(\mathbf{x}, \mathbf{x}', \tilde{\mathbf{x}}) f(\tilde{\mathbf{x}}) d\tilde{\mathbf{x}}. \quad (2)$$

where $f : \mathcal{A} \to \mathbb{R}$ denotes the *spatial loss field* (SLF) capturing the attenuation at location $\tilde{\mathbf{x}}$, and $w : \mathcal{A} \times \mathcal{A} \times \mathcal{A} \to \mathbb{R}$ is the weight function modeling the influence of the SLF at $\tilde{\mathbf{x}}$ to the shadowing experienced by link $\mathbf{x}$–$\mathbf{x}'$. Typically, $w$ confers a greater weight $w(\mathbf{x}, \mathbf{x}', \tilde{\mathbf{x}})$ to those locations $\tilde{\mathbf{x}}$ lying closer to the link $\mathbf{x}$–$\mathbf{x}'$. Examples of the weight function include the *normalized ellipse model* [30]

$$w(\mathbf{x}, \mathbf{x}', \tilde{\mathbf{x}}) := \begin{cases} 1/\sqrt{d(\mathbf{x}, \mathbf{x}')}, & \text{if } d(\mathbf{x}, \tilde{\mathbf{x}}) + d(\mathbf{x}', \tilde{\mathbf{x}}) \\ & \quad < d(\mathbf{x}, \mathbf{x}') + \lambda/2 \\ 0, & \text{otherwise} \end{cases} \quad (3)$$

where $\lambda > 0$ is a tunable parameter. The value of $\lambda$ is commonly set to the wavelength to assign non-zero weights only within the first Fresnel zone. In radio tomography, the integral in (2) is approximated as

$$s(\mathbf{x}, \mathbf{x}') \simeq c \sum_{i=1}^{N_g} w(\mathbf{x}, \mathbf{x}', \tilde{\mathbf{x}}_i) f(\tilde{\mathbf{x}}_i) \quad (4)$$

where $\{\tilde{\mathbf{x}}_i\}_{i=1}^{N_g}$ is a grid of points over $\mathcal{A}$ and $c$ is a constant that can be set to unity without loss of generality by absorbing any scaling factor in $f$. Clearly, (4) shows that $s(\mathbf{x}, \mathbf{x}')$ depends on $f$ only through its values at the grid points.

The model in (2) describes how the spatial distribution of obstructions in the propagation path influences the attenuation between a pair of locations. The usefulness of this model is twofold: i) as $f$ represents absorption across space, it can be used for imaging; and ii) once $f$ and $w$ are known, the gain between any two points $\mathbf{x}$ and $\mathbf{x}'$ can be recovered through (1) and (2), which is precisely the objective of channel-gain cartography.

The goal of radio tomography is to obtain a tomogram by estimating $f$. To this end, $N$ sensors located at $\{\mathbf{x}_1, \ldots, \mathbf{x}_N\} \in \mathcal{A}$ collaboratively obtain channel-gain measurements. At time slot $\tau$, the radios indexed by $n(\tau)$ and $n'(\tau)$ measure the channel-gain $\check{g}_\tau := g(\mathbf{x}_{n(\tau)}, \mathbf{x}_{n'(\tau)}) + \nu_\tau$ by exchanging training sequences known to both transmitting and receiving radios, where $n(\tau), n'(\tau) \in \{1, \ldots, N\}$ and $\nu_\tau$ denotes measurement noise. It is supposed that $g_0$ and $\gamma$ have been estimated during a calibration stage. After subtracting these from $\check{g}_\tau$, the shadowing estimate is found as

$$\begin{aligned} \check{s}_\tau &:= g_0 - \gamma 10 \log_{10} d(\mathbf{x}_{n(\tau)}, \mathbf{x}_{n'(\tau)}) - \check{g}_\tau \\ &= s(\mathbf{x}_{n(\tau)}, \mathbf{x}_{n'(\tau)}) - \nu_\tau. \end{aligned} \quad (5)$$

Having available $\check{\mathbf{s}}_t := [\check{s}_1, \ldots, \check{s}_t]^\top \in \mathbb{R}^t$ along with the known set of links $\{(\mathbf{x}_{n(\tau)}, \mathbf{x}_{n'(\tau)})\}_{\tau=1}^t$ and the weight function $w$, the problem is to estimate $f$, or equivalently $\boldsymbol{f} := [f(\tilde{\mathbf{x}}_1), \ldots, f(\tilde{\mathbf{x}}_{N_g})]^\top \in \mathbb{R}^{N_g}$ using (4).

Regularized least-squares (LS) estimators of $\boldsymbol{f}$ solve [9], [18], [30]

$$\min_{\boldsymbol{f}} \sum_{\tau=1}^t \left( \check{s}_\tau - \sum_{i=1}^{N_g} w(\mathbf{x}_{n(\tau)}, \mathbf{x}_{n'(\tau)}, \tilde{\mathbf{x}}_i) f(\tilde{\mathbf{x}}_i) \right)^2 + \mu_f \mathcal{R}(\boldsymbol{f}) \tag{6}$$

where $\mathcal{R} : \mathbb{R}^{N_g} \to \mathbb{R}$ is a generic regularizer to promote a known attribute of $\boldsymbol{f}$, and $\mu_f \geq 0$ is a regularization weight to reflect compliance of $\boldsymbol{f}$ with this attribute. Although (6) has been successfully applied to radio tomographic imaging tasks after customizing the regularizer to the propagation environment, how accurate approximation is provided by a regularized solution of (6) is unclear when the propagation environment exhibits inhomogeneous characteristics.

To overcome this and improve estimation performance of the SLF, a priori knowledge on the heterogeneous structure of $f$ will be exploited next, under a Bayesian framework.

## III. ADAPTIVE BAYESIAN RADIO TOMOGRAPHY

In this section, we view $\boldsymbol{f}$ as random, and forth propose a two-layer Bayesian SLF model, along with an MCMC-based approach for inference. We further develop an adaptive data acquisition strategy to select informative measurements.

### A. Bayesian model and problem formulation

Let $\mathcal{A}$ consist of two disjoint homogeneous regions $\mathcal{A}_0 := \{\mathbf{x} | \mathbb{E}[f(\mathbf{x})] = \mu_{f_0}, \text{Var}[f(\mathbf{x})] = \sigma_{f_0}^2, \mathbf{x} \in \mathcal{A}\}$, and $\mathcal{A}_1 := \{\mathbf{x} | \mathbb{E}[f(\mathbf{x})] = \mu_{f_1}, \text{Var}[f(\mathbf{x})] = \sigma_{f_1}^2, \mathbf{x} \in \mathcal{A}\}$, giving rise to a hidden label field $\boldsymbol{z} := [z(\tilde{\mathbf{x}}_1), \ldots, z(\tilde{\mathbf{x}}_{N_g})]^\top \in \{0, 1\}^{N_g}$ with binary entries $z(\tilde{\mathbf{x}}_i) = k$ if $\tilde{\mathbf{x}}_i \in \mathcal{A}_k$ $\forall i$, and $k = 0, 1$. The two separate regions can be used to model heterogeneous environments. For instance, if $\mathcal{A}$ corresponds to an urban area, $\mathcal{A}_1$ may include densely populated regions with buildings, while $\mathcal{A}_0$ with $\mu_{f_0} < \mu_{f_1}$ may capture the less obstructive open spaces. In such a paradigm, we model the conditional distribution of $f(\tilde{\mathbf{x}}_i)$ as

$$f(\tilde{\mathbf{x}}_i) | z(\tilde{\mathbf{x}}_i) = k \sim \mathcal{N}(\mu_{f_k}, \sigma_{f_k}^2), \tag{7}$$

while the Ising prior [28], which is a binary version of the discrete MRF Potts prior [12], is assigned to $\boldsymbol{z}$ in order to capture the dependency among spatially correlated labels. By the Hammersley-Clifford theorem [10], the Ising prior of $\boldsymbol{z}$ follows a Gibbs distribution

$$p(\boldsymbol{z}|\beta) = \frac{1}{C(\beta)} \exp \left[ \beta \sum_{i=1}^{N_g} \sum_{j \in \mathcal{N}(\tilde{\mathbf{x}}_i)} \delta(z(\tilde{\mathbf{x}}_j) = z(\tilde{\mathbf{x}}_i)) \right] \tag{8}$$

where $\mathcal{N}(\tilde{\mathbf{x}}_i)$ is a set of indices associated with 1-hop neighbors of $\tilde{\mathbf{x}}_i$ on the rectangular grid in Fig. 1, $\beta$ is a granularity coefficient controlling the degree of homogeneity in $\boldsymbol{z}$, $\delta(\cdot)$ is Kronecker's delta, and

$$C(\beta) := \sum_{\boldsymbol{z} \in \mathcal{Z}} \exp \left[ \beta \sum_{i=1}^{N_g} \sum_{j \in \mathcal{N}(\tilde{\mathbf{x}}_i)} \delta(z(\tilde{\mathbf{x}}_j) = z(\tilde{\mathbf{x}}_i)) \right] \tag{9}$$

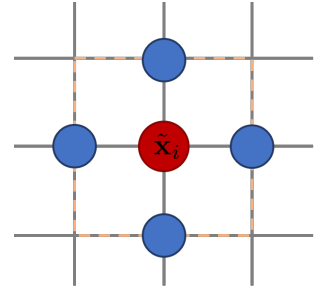

Fig. 1: Four-connected MRF with $z(\tilde{\mathbf{x}}_i)$ marked red and its neighbors in $\mathcal{N}(\tilde{\mathbf{x}}_i)$ marked blue.

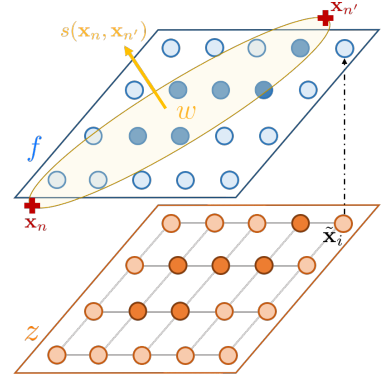

Fig. 2: Gauss-Potts model for radio tomography, together with the measurement model for sensors located at $(\mathbf{x}_n, \mathbf{x}_{n'})$.

is the partition function with $\mathcal{Z} := \{0, 1\}^{N_g}$. By assuming conditional independence of $\{f(\tilde{\mathbf{x}}_i)\}_{i=1}^{N_g}$ given $\boldsymbol{z}$, the resulting model is referred to as the Gauss-Potts model [2] with two labels. The Gauss-Potts model for radio tomography is depicted in Fig. 2 with the measurement model in (2).

To describe priors of other parameters, let $\nu_t$ be independent and identically distributed (i.i.d) Gaussian with zero mean and variance $\sigma_\nu^2$, and $\boldsymbol{\theta}$ denote a hyperparameter vector comprising $\sigma_\nu^2$, $\beta$, and $\boldsymbol{\theta}_f := [\mu_{f_0}, \mu_{f_1}, \sigma_{f_0}^2, \sigma_{f_1}^2]^\top$. The weight matrix $\mathbf{W} \in \mathbb{R}^{N_g \times t}$ is formed with columns $\mathbf{w}_\tau := [w(\mathbf{x}_{n(\tau)}, \mathbf{x}_{n'(\tau)}, \tilde{\mathbf{x}}_1), \ldots, w(\mathbf{x}_{n(\tau)}, \mathbf{x}_{n'(\tau)}, \tilde{\mathbf{x}}_{N_g})]^\top \in \mathbb{R}^{N_g}$ of the link $\mathbf{x}_{n(\tau)} - \mathbf{x}_{n'(\tau)}$ for $\tau = 1, \ldots, t$. Assuming the independence among entries of $\boldsymbol{\theta}$, $p(\boldsymbol{\theta})$ can be expressed as

$$p(\boldsymbol{\theta}) = p(\sigma_\nu^2) p(\beta) p(\boldsymbol{\mu}_{f_k}) p(\boldsymbol{\sigma}_{f_k}^2) \tag{10}$$

with $p(\boldsymbol{\mu}_{f_k}) = p(\mu_{f_0}) p(\mu_{f_1})$ and $p(\boldsymbol{\sigma}_{f_k}^2) = p(\sigma_{f_0}^2) p(\sigma_{f_1}^2)$, where the individual priors $p(\sigma_\nu^2)$, $p(\beta)$, $p(\boldsymbol{\mu}_{f_k})$, and $p(\boldsymbol{\sigma}_{f_k}^2)$ are specified next.

*1) Granularity coefficient $\beta$:* To cope with the variability of $\beta$ in accordance with structural patterns of the propagation medium, $\beta$ is viewed as an unknown random variable that is to be estimated together with $\boldsymbol{f}$ and $\boldsymbol{z}$ under the Bayesian framework. Similar to e.g., [25], the uniform distribution is adopted for the prior of $\beta$ as

$$p(\beta) = \mathcal{U}_{(0, \beta_{\max})}(\beta) := \begin{cases} 1/\beta_{\max}, & \text{if } \beta \in [0, \beta_{\max}] \\ 0, & \text{otherwise.} \end{cases} \tag{11}$$

*2) Noise variance $\sigma_\nu^2$:* In the presence of the additive Gaussian noise with fixed mean, it is common to assign a

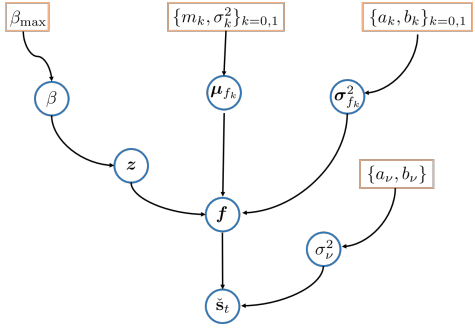

Fig. 3: Graphical representation of the hierarchical Bayesian model for (hyper) parameters (those in boxes are fixed).

conjugate prior to $\sigma_\nu^2$, which reproduces a posterior distribution in the same family of its prior. The inverse gamma (IG) distribution serves this purpose for $\sigma_\nu^2 \in \mathbb{R}^+$ as follows:

$$p(\sigma_\nu^2) = \mathcal{IG}(a_\nu, b_\nu) := \frac{b_\nu^{a_\nu}}{\Gamma(a_\nu)} (\sigma_\nu^2)^{-a_\nu - 1} \exp\left(-\frac{b_\nu}{\sigma_\nu^2}\right) \quad (12)$$

where $a_\nu$ is referred to as the shape parameter, $b_\nu$ as the scale parameter, and $\Gamma(\cdot)$ denotes the gamma function.

*3) Hyperparameters of the SLF $\boldsymbol{\theta}_f$:* While the prior for $\mu_{f_k}$ is assumed to be Gaussian with mean $m_k$ and variance $\sigma_k^2$ (see also [2]), the inverse Gamma distribution parameterized by $\{a_k, b_k\}$ is considered for the prior of $\sigma_{f_k}^2$:

$$p(\mu_{f_k}) = \mathcal{N}(m_k, \sigma_k^2), \quad k = 0, 1, \quad (13)$$
$$p(\sigma_{f_k}^2) = \mathcal{IG}(a_k, b_k), \quad k = 0, 1. \quad (14)$$

Together with the priors for $\{\boldsymbol{f}, \boldsymbol{z}, \boldsymbol{\theta}\}$, our joint posterior is

$$p(\boldsymbol{f}, \boldsymbol{z}, \boldsymbol{\theta} | \check{\boldsymbol{s}}_t) \propto p(\check{\boldsymbol{s}}_t | \boldsymbol{f}, \sigma_\nu^2) p(\boldsymbol{f} | \boldsymbol{z}, \boldsymbol{\theta}_f) p(\boldsymbol{z} | \beta) p(\boldsymbol{\theta}) \quad (15)$$

where $p(\check{\boldsymbol{s}}_t | \boldsymbol{f}, \sigma_\nu^2)$ is the data likelihood. Note that Fig. 3 summarizes the proposed hierarchical Bayesian model for $\{\check{\boldsymbol{s}}_t, \boldsymbol{f}, \boldsymbol{z}, \boldsymbol{\theta}\}$ as a directed acyclic graph, where the dependency between (hyper) parameters is indicated with an arrow.

We will pursue the the conditional MMSE estimator

$$\hat{\boldsymbol{f}}_{\text{MMSE}} := \mathbb{E}[\boldsymbol{f} | \boldsymbol{z} = \hat{\boldsymbol{z}}_{\text{MAP}}, \check{\boldsymbol{s}}_t] \quad (16)$$

where the marginal MAP estimate is

$$\hat{\boldsymbol{z}}_{\text{MAP}} := \arg\max_{\boldsymbol{z}} p(\boldsymbol{z} | \check{\boldsymbol{s}}_t). \quad (17)$$

Furthermore, the marginal MMSE estimates of the $\boldsymbol{\theta}$ entries are found as

$$\widehat{\sigma_\nu^2}_{\text{MMSE}} := \mathbb{E}[\sigma_\nu^2 | \check{\boldsymbol{s}}_t] \quad (18)$$
$$\hat{\beta}_{\text{MMSE}} := \mathbb{E}[\beta | \check{\boldsymbol{s}}_t] \quad (19)$$
$$\widehat{\mu_{f_k}}_{\text{MMSE}} := \mathbb{E}[m_k | \check{\boldsymbol{s}}_t], \quad k = 0, 1 \quad (20)$$
$$\widehat{\sigma_{f_k}^2}_{\text{MMSE}} := \mathbb{E}[\sigma_k^2 | \check{\boldsymbol{s}}_t], \quad k = 0, 1. \quad (21)$$

### B. Sampling via Markov chain Monte Carlo

While approximate estimators have been proposed for Bayesian inference (see e.g., [13], [35]), analytical solutions to (16)–(21) are not tractable due to the complex form of the posterior in (15) that does not permit marginalization or maximization. To bypass this challenge, one can generate samples from (15), and then numerically approximate the desired estimators from those samples. MCMC is a class of methods used to generate samples from a complex distribution [8].

Among MCMC methods, Gibbs sampling [7] is particularly suitable for this work. It draws samples following the target distribution (e.g., the posterior in (15)) by sweeping through each variable to sample from its conditional distribution while fixing the others to their up-to-date values. Although the samples at early iterations of Gibbs sampling with random initialization are not representative of the desired distribution (such duration is called the *burn-in* period $N_{\text{Burn-in}}$), the theory of MCMC guarantees that the stationary distribution of those samples matches with the target distribution [8].

Gibbs sampling requires only the conditional distribution within a proportionality scale. When a given conditional distribution is not easy to simulate, one can resort to a Metropolis-Hastings (MH) sampler [11], which generates a candidate from a simple proposal distribution of such conditional distribution, and accepts (or rejects) the candidate as a sample of interest under a certain acceptance ratio $\alpha$. The substitution of MH sampling for some sampling steps inside the Gibbs sampler results in a Metropolis-within-Gibbs (MwG) sampler, as listed in Alg. 1. Posterior conditionals considered in this work and associated sampling methods will be described next.

*1) Spatial loss field $\boldsymbol{f}$:* It is easy to show that

$$p(\boldsymbol{f} | \check{\boldsymbol{s}}_t, \boldsymbol{z}, \boldsymbol{\theta}) \propto p(\check{\boldsymbol{s}}_t | \boldsymbol{f}, \sigma_\nu^2) p(\boldsymbol{f} | \boldsymbol{z}, \boldsymbol{\theta}_f)$$
$$\sim \mathcal{N}(\check{\boldsymbol{\mu}}_{f|z,\boldsymbol{\theta},\check{\boldsymbol{s}}_t}, \boldsymbol{\Sigma}_{f|z,\boldsymbol{\theta},\check{\boldsymbol{s}}_t}) \quad (22)$$

where

$$\boldsymbol{\Sigma}_{f|z,\boldsymbol{\theta},\check{\boldsymbol{s}}_t} := \left((\sigma_\nu^2)^{-1} \boldsymbol{W}\boldsymbol{W}^\top + \boldsymbol{\Delta}_{f|z}^{-1}\right)^{-1} \quad (23)$$

$$\check{\boldsymbol{\mu}}_{f|z,\boldsymbol{\theta},\check{\boldsymbol{s}}_t} := \boldsymbol{\Sigma}_{f|z,\boldsymbol{\theta},\check{\boldsymbol{s}}_t} \left((\sigma_\nu^2)^{-1} \boldsymbol{W}\check{\boldsymbol{s}}_t + \boldsymbol{\Delta}_{f|z}^{-1} \boldsymbol{\mu}_{f|z}\right) \quad (24)$$

since $p(\boldsymbol{f}|\boldsymbol{z}, \boldsymbol{\theta}_f)$ follows $\mathcal{N}(\boldsymbol{\mu}_{f|z}, \boldsymbol{\Delta}_{f|z})$ by (7), with $\boldsymbol{\mu}_{f|z} := \mathbb{E}[\boldsymbol{f}|\boldsymbol{z}]$ and $\boldsymbol{\Delta}_{f|z} := \text{diag}(\{\text{Var}[f_i|z_i]\}_{i=1}^{N_g})$ with $f_i := f(\tilde{\boldsymbol{x}}_i)$ and $z_i := z(\tilde{\boldsymbol{x}}_i)$ (see Appendix A for derivation). Hence, $\boldsymbol{f}$ can be easily simulated by a standard sampling method.

---

**Algorithm 1** Metropolis-within-Gibbs sampler for $\{\boldsymbol{f}, \boldsymbol{z}, \boldsymbol{\theta}\}$

**Input:** $\boldsymbol{z}^{(0)}, \boldsymbol{\theta}^{(0)}, \check{\boldsymbol{s}}_t, N_{\text{CL}}, N_{\text{Burn-in}}$, and $N_{\text{Iter}}$
1: **for** $l = 1$ to $N_{\text{Iter}}$ **do**
2:     Generate $\boldsymbol{f}^{(l)} \sim p(\boldsymbol{f}|\check{\boldsymbol{s}}_t, \boldsymbol{z}^{(l-1)}, \boldsymbol{\theta}^{(l-1)})$ in (22)
3:     Generate $\boldsymbol{z}^{(l)} \sim p(\boldsymbol{z}|\check{\boldsymbol{s}}_t, \boldsymbol{f}^{(l)}, \boldsymbol{\theta}^{(l-1)})$ via Alg. 2
4:     Generate $\beta^{(l)} \sim p(\beta|\check{\boldsymbol{s}}_t, \boldsymbol{f}^{(l)}, \boldsymbol{z}^{(l)}, \sigma_\nu^{2(l-1)}, \boldsymbol{\theta}_f^{(l-1)})$ via Alg. 4
5:     Generate $\sigma_\nu^{2(l)} \sim p(\sigma_\nu^2|\check{\boldsymbol{s}}_t, \boldsymbol{f}^{(l)}, \boldsymbol{z}^{(l)}, \beta^{(l)}, \boldsymbol{\theta}_f^{(l-1)})$ in (32)
6:     Generate $\mu_{f_k}^{(l)} \sim p(\mu_{f_k}|\check{\boldsymbol{s}}_t, \boldsymbol{f}^{(l)}, \boldsymbol{z}^{(l)}, \sigma_\nu^{2(l)}, \beta^{(l)}, \boldsymbol{\sigma}_{f_k}^{2(l-1)})$ in (34) for $k = 0, 1$
7:     Generate $\sigma_{f_k}^{2(l)} \sim p(\sigma_{f_k}^2|\check{\boldsymbol{s}}_t, \boldsymbol{f}^{(l)}, \boldsymbol{z}^{(l)}, \sigma_\nu^{2(l)}, \beta^{(l)}, \boldsymbol{\mu}_{f_k}^{(l)})$ in (38) for $k = 0, 1$
8: **end for**
9: **return** $\mathcal{S}^{(t)} := \{\boldsymbol{f}^{(l)}, \boldsymbol{z}^{(l)}, \boldsymbol{\theta}^{(l)}\}_{l=N_{\text{Burn-in}}+1}^{N_{\text{Iter}}}$

**Algorithm 2** Single-site Gibbs sampler for $z$

**Input:** $f^{(l)}$ and $z^{(l-1)}$
1: Initialize $\zeta^{(l)} := [\zeta_1^{(l)}, \ldots, \zeta_{N_g}^{(l)}]^\top = z^{(l-1)}$
2: **for** $i = 1$ to $N_g$ **do**
3:     Obtain $h_i$ in (26) with $z = \zeta^{(l)}$ and $f = f^{(l)}$
4:     Generate $u \sim \mathcal{U}_{(0,1)}$
5:     **if** $u < (1 + h_i)^{-1}$ **then**
6:        Set $\zeta_i^{(l)} = 1$
7:     **else**
8:        Set $\zeta_i^{(l)} = 0$
9:     **end if**
10: **end for**
11: **return** $z^{(l)} = \zeta^{(l)}$

**Algorithm 3** Single-site Gibbs sampler for $z^*$

**Input:** $z^{(l)}$, $\beta^*$, and $N_{\text{CL}}$
1: Initialize $\zeta^* := [\zeta_1^*, \ldots, \zeta_{N_g}^*]^\top = z^{(l)}$
2: **for** $m = 1$ to $N_{\text{CL}}$ **do**
3:     **for** $i = 1$ to $N_g$ **do**
4:        Obtain $h_i^*$ in (29) with $z^* = \zeta^*$
5:        Generate $u \sim \mathcal{U}_{(0,1)}$
6:        **if** $u < (1 + h_i^*)^{-1}$ **then**
7:           Set $\zeta_i^* = 1$
8:        **else**
9:           Set $\zeta_i^* = 0$
10:        **end if**
11:     **end for**
12: **end for**
13: **return** $z^* = \zeta^*$

*2) Hidden label field $z$:* A Gibbs sampler is required to simulate $p(z|\check{s}_t, f, \theta) \propto p(f|z, \theta_f)p(z|\beta)$ while avoiding the intractable computation of $C(\beta)$ in (8). Let $z_{-i}$ and $z_{\mathcal{N}(\tilde{x}_i)}$ represent replicas of $z$ without its $i$-th entry, and only with the entries of $\mathcal{N}(\tilde{x}_i)$, respectively. By the Markovianity of $z$ and conditional independence between $f_i$ and $f_j$ $\forall i \neq j$ given $z$, the conditional distribution of $z_i$ is

$$p(z_i|z_{-i}, \check{s}_t, f, \theta) \propto \exp\left[\ell(z_i) + \beta \sum_{j \in \mathcal{N}(\tilde{x}_i)} \delta(z_j = z_i)\right] \quad (25)$$

where $\ell(z_i) := \ln p(f_i|z_i, \theta_f)$. After evaluating (25) for $z_i = 0, 1$ and normalizing, one can obtain $p(z_i = 1|z_{-i}, \check{s}_t, f, \theta) = (1 + h_i)^{-1}$, where

$$h_i := \exp\left[\ell(z_i = 0) - \ell(z_i = 1) + \sum_{j \in \mathcal{N}(\tilde{x}_i)} \beta(1 - 2z_j)\right] \quad (26)$$

with $\delta(z_j = 0) - \delta(z_j = 1) = 1 - 2z_j$. Then, the sample of $z$ can be obtained via the single-site Gibbs sampler by using (26), as summarized in Alg. 2. It is worth stressing that the sampling criterion with $h_i$ in (26) does not require the evaluation of $C(\beta)$.

*3) Granularity coefficient $\beta$:* The conditional distribution of $\beta$ satisfies the following proportionality relation

$$p(\beta|\check{s}_t, f, z, \sigma_\nu^2, \theta_f) \propto p(z|\beta)p(\beta) \quad (27)$$
$$\propto \frac{1}{\beta_{\max} C(\beta)} \exp\left[\beta \sum_{i=1}^{N_g} \sum_{j \in \mathcal{N}(\tilde{x}_i)} \delta(z_j = z_i)\right]$$

for $\beta \in [0, \beta_{\max}]$, simply by the Gibbs distribution in (8) and the uniform prior of $\beta$ in (11). Unfortunately, sampling of $\beta$ is formidably challenging because evaluating the partition function $C(\beta)$ in $p(z|\beta)$, incurs exponential complexity. To address this, one may resort to auxiliary variable MCMC methods that do not require exact evaluation of $p(z|\beta)$, including the single auxiliary variable method (SAVM) [22] and the exchange algorithm [23]. Those methods replace $C(\beta)$ with its single-point importance sampling estimate by using an auxiliary variable, which unfortunately must be generated via exact sampling that is generally expensive for statistical models with intractable partition functions. To bypass exact sampling for generating this auxiliary variable, we will leverage a *double-MH* sampling method for $\beta$; also [19].

Let $z^*$ and $\beta^*$ denote the auxiliary variable of $z$ and a candidate of $\beta$ for MH sampling, respectively. The idea behind the double-MH algorithm is to generate $z^*$ through $N_{\text{CL}}$ cycles of MH updates from the current sample $z^{(l)}$, instead of using exact sampling from $p(z^*|\beta^*)$. As the name suggests, the double-MH sampling includes two nested MCMC samplers: the inner one to generate a chain of the auxiliary variable at each step of the outer sampler for $\beta$. It is instructive to mention that $N_{\text{CL}}$ is not necessarily large by initializing the chain with $z^{(l)}$ at the $l$-th iteration [25], [19], which means that additional complexity to generate the auxiliary variable is not necessarily high. In this work, $z^*$ is obtained via another single-site Gibbs sampler, as described in Alg. 3:

$$p(z_i^*|z_{-i}^*, \beta^*) \propto \exp\left[\beta^* \sum_{j \in \mathcal{N}(\tilde{x}_i)} \delta(z_j^* = z_i^*)\right] \quad \forall i \quad (28)$$

and a sample of $z_i^*$ is generated by utilizing $p(z_i^* = 1|z_{-i}^*, \beta^*) = (1 + h_i^*)^{-1}$ with

$$h_i^* := \exp\left[\sum_{j \in \mathcal{N}(\tilde{x}_i)} \beta(1 - 2z_j^*)\right]. \quad (29)$$

The overall double-MH sampler for $\beta$ is summarized in Alg. 4. A proposal distribution of $\beta^*$ is the truncated Gaussian

$$q(\beta^*|\beta^{(l-1)}) = \begin{cases} \mathcal{N}(\beta^{(l-1)}, \sigma_q^2)/c, & \text{if } \beta^* \in [0, \beta_{\max}] \\ 0, & \text{otherwise} \end{cases} \quad (30)$$

with a tunable variable $\sigma_q^2$ and a normalizing constant

$$c := \int_0^{\beta_{\max}} \frac{1}{\sqrt{2\pi\sigma_q^2}} \exp\left[-\frac{1}{2\sigma_q^2}(\beta^* - \beta^{(l-1)})^2\right] d\beta^*. \quad (31)$$

*4) Noise variance $\sigma_\nu^2$:* With $p(\sigma_\nu^2)$ in (12), we have the posterior conditional of $\sigma_\nu^2$ satisfying

$$p(\sigma_\nu^2|\check{s}_t, f, z, \beta, \theta_f) \propto p(\check{s}_t|f, \sigma_\nu^2)p(\sigma_\nu^2)$$
$$\propto \mathcal{IG}(a_\nu + \frac{t}{2}, b_\nu + \frac{1}{2}\|\check{s}_t - \mathbf{W}^\top f\|_2^2). \quad (32)$$

**Algorithm 4** Double-MH sampler for $\beta$

**Input:** $\beta^{(l-1)}$, $z^{(l)}$, and $N_{\text{CL}}$
1: Generate $\beta^* \sim q(\beta^*|\beta^{(l-1)})$ in (30)
2: Generate $z^* \sim p(z^*|\beta^*)$ via Alg. 3
3: Set $\alpha' := \frac{p(\beta^*)q(\beta^{(l-1)}|\beta^*)p(z^*|\beta^{(l-1)})p(z^{(l)}|\beta^*)}{p(\beta^{(l-1)})q(\beta^*|\beta^{(l-1)})p(z^{(l)}|\beta^{(l-1)})p(z^*|\beta^*)}$
4: Obtain $\alpha = \min\{1, \alpha'\}$
5: Generate $u \sim \mathcal{U}_{(0,1)}$
6: **if** $u < \alpha$ **then**
7: $\quad \beta^{(l)} = \beta^*$
8: **else**
9: $\quad \beta^{(l)} = \beta^{(l-1)}$
10: **end if**
11: **return** $\beta^{(l)}$

Therefore, a sample of $\sigma_\nu^2$ can be generated by a standard sampling method.

*5) Means of the SLF $\boldsymbol{\mu}_{f_k}$:* Let $\boldsymbol{f}_k$ be the $N_k \times 1$ vector formed by concatenating $f(\tilde{\mathbf{x}}_i)$ for $\tilde{\mathbf{x}}_i \in \mathcal{A}_k$, for $k = 0, 1$. By recalling the priori independence between the parameters of disjoint homogeneous regions $\mathcal{A}_0$ and $\mathcal{A}_1$, the posterior conditional of $\boldsymbol{\mu}_{f_k} := [\mu_{f_0}, \mu_{f_1}]^\top$ can be expressed as

$$p(\boldsymbol{\mu}_{f_k}|\check{\mathbf{s}}_t, \boldsymbol{f}, \boldsymbol{z}, \sigma_\nu^2, \beta, \sigma_{f_0}^2, \sigma_{f_1}^2) \quad (33)$$
$$\propto p(\boldsymbol{f}|\boldsymbol{z}, \boldsymbol{\theta}_f)p(\boldsymbol{\mu}_{f_k}) \propto p(\mu_{f_0}|\boldsymbol{z}, \boldsymbol{f}_0, \sigma_{f_0}^2)p(\mu_{f_1}|\boldsymbol{z}, \boldsymbol{f}_1, \sigma_{f_1}^2)$$

with

$$p(\mu_{f_k}|\boldsymbol{z}, \boldsymbol{f}_k, \sigma_{f_k}^2) \propto p(\boldsymbol{f}_k|\boldsymbol{z}, \mu_{f_k}, \sigma_{f_k}^2)p(\mu_{f_k}), \ \forall k. \quad (34)$$

Since a sample of each $\mu_{f_k}$ can be independently drawn according to $p(\mu_{f_k}|\boldsymbol{z}, \boldsymbol{f}_k, \sigma_{f_k}^2)$ in (34), the sampling method for $\mu_{f_k}$ will be described.

To efficiently simulate a sample of $\mu_{f_k}$, the likelihood $p(\boldsymbol{f}_k|\boldsymbol{z}, \mu_{f_k}, \sigma_{f_k}^2)$ is recast as an univariate distribution with respect to the sample mean $\bar{f}_k := (\sum_i f_{k,i})/N_k$ as

$$p(\boldsymbol{f}_k|\boldsymbol{z}, \mu_{f_k}, \sigma_{f_k}^2) \propto \exp\left[-\frac{1}{2\sigma_{f_k}^2} \sum_{i=1}^{N_k} (f_{k,i} - \mu_{f_k})^2\right]$$
$$\propto \exp\left[-\frac{1}{2\sigma_{f_k}^2}(-2\mu_{f_k}\sum_{i=1}^{N_k} f_{k,i} + N_k \mu_{f_k}^2)\right]$$
$$\propto \exp\left[-\frac{N_k}{2\sigma_{f_k}^2}(\bar{f}_k - \mu_{f_k})^2\right]$$
$$\propto \mathcal{N}(\mu_{f_k}, 2\sigma_{f_k}^2/N_k). \quad (35)$$

Since $p(\mu_{f_k})$ is the Gaussian conjugate prior, one can show that $p(\mu_{f_k}|\boldsymbol{z}, \boldsymbol{f}_k, \sigma_{f_k}^2)$ is Gaussian as well, parameterized by

$$\mathbb{E}\left[\mu_{f_k}|\boldsymbol{z}, \boldsymbol{f}_k, \sigma_{f_k}^2\right] = \frac{\sigma_k^2 \bar{f}_k}{\sigma_k^2 + (\sigma_{f_k}^2/N_k)} + \frac{\sigma_{f_k}^2/N_k}{\sigma_k^2 + (\sigma_{f_k}^2/N_k)}m_k$$

$$\text{Var}\left[\mu_{f_k}|\boldsymbol{z}, \boldsymbol{f}_k, \sigma_{f_k}^2\right] = \left(\frac{1}{\sigma_k^2} + \frac{N_k}{\sigma_{f_k}^2}\right)^{-1}. \quad (36)$$

Therefore, a sample of $\mu_{f_k}$ can be generated for $k = 0, 1$ by using a standard sampling method.

*6) Variances of the SLF $\boldsymbol{\sigma}_{f_k}^2$:* Similar to $\boldsymbol{\mu}_{f_k}$, the statistical independence between $\mathcal{A}_0$ and $\mathcal{A}_1$ leads to the following proportionality of the posterior conditional for $\boldsymbol{\sigma}_{f_k}^2 := [\sigma_{f_0}^2, \sigma_{f_1}^2]^\top$

$$p(\boldsymbol{\sigma}_{f_k}^2|\check{\mathbf{s}}_t, \boldsymbol{f}, \boldsymbol{z}, \sigma_\nu^2, \beta, \boldsymbol{\mu}_{f_k}) \propto p(\boldsymbol{f}|\boldsymbol{z}, \boldsymbol{\theta}_f)p(\boldsymbol{\sigma}_{f_k}^2)$$
$$\propto p(\sigma_{f_0}^2|\boldsymbol{z}, \boldsymbol{f}_0, \mu_{f_0})p(\sigma_{f_1}^2|\boldsymbol{z}, \boldsymbol{f}_1, \mu_{f_1}) \quad (37)$$

where

$$p(\sigma_{f_k}^2|\boldsymbol{z}, \boldsymbol{f}_k, \mu_{f_k}) \propto p(\boldsymbol{f}_k|\boldsymbol{z}, \mu_{f_k}, \sigma_{f_k}^2)p(\sigma_{f_k}^2)$$
$$\propto \mathcal{IG}(a_k + \frac{N_k}{2}, b_k + \frac{1}{2}\|\boldsymbol{f}_k - \mu_{f_k}\mathbf{1}_{N_k}\|_2^2), \ \forall k. \quad (38)$$

Therefore, a sample of each $\sigma_k^2$ can be independently drawn according to $p(\sigma_{f_k}^2|\boldsymbol{z}, \boldsymbol{f}_k, \mu_{f_k})$ in (38).

### C. Efficient estimators for $\boldsymbol{f}$, $\boldsymbol{z}$, and $\boldsymbol{\theta}$.

In this section, efficient sample-based estimators for $\boldsymbol{f}$, $\boldsymbol{z}$, and $\boldsymbol{\theta}$ are derived, by using a set of samples $\mathcal{S}^{(t)}$ from Alg. 1. Building on [13], the elementwise marginal MAP estimator of $\boldsymbol{z}$ and its sample-based approximation are

$$\hat{z}_{i,\text{MAP}} = \arg\max_{z_i \in \{0,1\}} p(z_i|\check{\mathbf{s}}_t)$$
$$\simeq \arg\max_{z_i \in \{0,1\}} \frac{1}{|\mathcal{S}^{(t)}|} \sum_{l=N_{\text{Burn-in}}+1}^{N_{\text{Iter}}} \delta(z_i^{(l)} = z_i) \quad (39)$$

for $i = 1, \ldots, N_g$. After obtaining $\hat{z}_{\text{MAP}}$, the sample-based elementwise conditional MMSE estimator of $\boldsymbol{f}$ follows as

$$\hat{f}_{i,\text{MMSE}} \simeq \frac{1}{|\mathcal{S}_i^{(t)}|} \sum_{l=N_{\text{Burn-in}}+1}^{N_{\text{Iter}}} f_i^{(l)} \delta(z_i^{(l)} = \hat{z}_{i,\text{MAP}}), \quad \forall i \quad (40)$$

where $\mathcal{S}_i^{(t)} \subset \mathcal{S}^{(t)}$ is a subset of samples such that $z_i^{(l)} = \hat{z}_{i,\text{MAP}}$ for $l = N_{\text{Burn-in}} + 1, \ldots, N_{\text{Iter}}$. To estimate $\boldsymbol{\theta}$, the following marginal MMSE estimators are employed

$$\hat{\beta}_{\text{MMSE}} \simeq \frac{1}{|\mathcal{S}^{(t)}|} \sum_{l=N_{\text{Burn-in}}+1}^{N_{\text{Iter}}} \beta^{(l)} \quad (41)$$

$$\widehat{\sigma_\nu^2}_{\text{MMSE}} \simeq \frac{1}{|\mathcal{S}^{(t)}|} \sum_{l=N_{\text{Burn-in}}+1}^{N_{\text{Iter}}} \sigma_\nu^{2(l)} \quad (42)$$

$$\widehat{\mu_{f_k}}_{\text{MMSE}} \simeq \frac{1}{|\mathcal{S}^{(t)}|} \sum_{l=N_{\text{Burn-in}}+1}^{N_{\text{Iter}}} \mu_{f_k}^{(l)}, \ k = 0, 1 \quad (43)$$

$$\widehat{\sigma_{f_k}^2}_{\text{MMSE}} \simeq \frac{1}{|\mathcal{S}^{(t)}|} \sum_{l=N_{\text{Burn-in}}+1}^{N_{\text{Iter}}} \sigma_{f_k}^{2(l)}, \ k = 0, 1. \quad (44)$$

**Remark 1 (Monitoring sampler-convergence).** The proposed sampler in Alg. 1 generates a sequence of samples from the desired distribution in (15), after a burn-in period to diminish the influence of initialization. By recalling that the stationary distribution of those samples is matched with the desired distribution, monitoring convergence of sample-sequences guides the choice of $N_{\text{Burn-in}}$.

Let $\psi$ denote a generic scalar random variable of interest. Suppose that $N_{\text{Seq}}$ parallel sequences of length $N_{\text{Iter}}$ are available, and let $\psi^{(l,m)}$ denote the $l$-th sample of $\psi$ in the $m$-th sequence for $l = 1, \ldots, N_{\text{Iter}}$ and $m = 1, \ldots, N_{\text{Seq}}$. Then, the following potential scale reduction factor (PSRF) estimate is adopted for convergence diagnosis [8]

$$\text{PSRF}(\psi) := \frac{N'_{\text{Iter}} - 1}{N'_{\text{Iter}}} + \frac{\sigma^2_{\text{Between}}}{\sigma^2_{\text{Within}}} \quad (45)$$

where $N'_{\text{Iter}} := N_{\text{Iter}} - N_{\text{Burn-in}}$, the within-sequence variance:

$$\sigma^2_{\text{Within}} := \frac{1}{N_{\text{Seq}}} \sum_{m=1}^{N_{\text{Seq}}} \frac{1}{N'_{\text{Iter}} - 1} \sum_{l=N_{\text{Burn-in}}+1}^{N_{\text{Iter}}} \left(\psi^{(l,m)} - \bar{\psi}^{(m)}\right)^2 \quad (46)$$

with $\bar{\psi}^{(m)} := \sum_{l=N_{\text{Burn-in}}+1}^{N_{\text{Iter}}} \psi^{(l,m)}/(N'_{\text{Iter}} - 1) \ \forall m$, and the between-sequence variance:

$$\sigma^2_{\text{Between}} := \frac{1}{N_{\text{Seq}}} \sum_{m=1}^{N_{\text{Seq}}} \left(\bar{\psi}^{(m)} - \bar{\psi}\right)^2 \quad (47)$$

with $\bar{\psi}^{(m)} := \sum_{m=1}^{N_{\text{Seq}}} \bar{\psi}^{(m)}/N_{\text{Seq}}$. As those sequences converge while $N_{\text{Iter}} \to \infty$, the PSRF declines to 1. In practice, each sequence is supposed to follow the desired distribution when PSRF $\leq 1.2$ [8, p. 138]. For synthetic data tests, $N_{\text{Burn-in}}$ and $N_{\text{Iter}}$ were found to have PSRF $\leq 1.06$ for $\boldsymbol{f}$, $\boldsymbol{z}$, and $\boldsymbol{\theta}$ over $N_{\text{Seq}} = 20$ independent sequences. On the other hand, $N_{\text{Burn-in}}$ and $N_{\text{Iter}}$ for real data tests were found to have PSRF $\leq 1.04$ for $\boldsymbol{f}$ and $\boldsymbol{z}$, while the PSRF $< 1.5$ for $\boldsymbol{\theta}$, over $N_{\text{Seq}} = 20$ independent sequences. It allows to have moderate-sized $N_{\text{Burn-in}}$ and $N_{\text{Iter}}$ for real data tests. Note that elementwise $\{\text{PSNR}(f_i), \text{PSNR}(z_i)\}_{i=1}^{N_g}$ were monitored for $\boldsymbol{f}$ and $\boldsymbol{z}$.

### D. Adaptive data acquisition via uncertainty sampling

The proposed Bayesian radio tomography accounts for the uncertainty of $\boldsymbol{f}$, through the variance in (23). Using the latter, our idea is to adaptively collect a measurement (or a mini-batch of measurements) from the set of available sensing radio pairs, with the goal of reducing the uncertainty of $\boldsymbol{f}$. To this end, we will rely on the conditional entropy [3] that in our context is given by

$$H_\tau(\boldsymbol{f}|\check{\boldsymbol{s}}_\tau, \boldsymbol{z}, \boldsymbol{\theta}) = \sum_{\boldsymbol{z}' \in \mathcal{Z}} \int_{\boldsymbol{\theta}', \check{\boldsymbol{s}}'_\tau} p(\check{\boldsymbol{s}}'_\tau, \boldsymbol{z}', \boldsymbol{\theta}') \quad (48)$$
$$\times H_\tau(\boldsymbol{f}|\check{\boldsymbol{s}}_\tau = \check{\boldsymbol{s}}'_\tau, \boldsymbol{z} = \boldsymbol{z}', \boldsymbol{\theta} = \boldsymbol{\theta}')d\boldsymbol{\theta}' d\check{\boldsymbol{s}}'_\tau$$

where

$$H_\tau(\boldsymbol{f}|\check{\boldsymbol{s}}_\tau = \check{\boldsymbol{s}}'_\tau, \boldsymbol{z} = \boldsymbol{z}', \boldsymbol{\theta} = \boldsymbol{\theta}')$$
$$:= -\int p(\boldsymbol{f}|\check{\boldsymbol{s}}'_\tau, \boldsymbol{z}', \boldsymbol{\theta}') \ln p(\boldsymbol{f}|\check{\boldsymbol{s}}'_\tau, \boldsymbol{z}', \boldsymbol{\theta}')d\boldsymbol{f}$$
$$= \frac{1}{2} \ln(|\boldsymbol{\Sigma}_{f|z'}|) + \frac{N_g}{2}\left(1 + \ln(2\pi)\right) \quad (49)$$

and $|\cdot|$ denotes matrix determinant. To obtain $\check{s}_{\tau+1}$, one can choose a pair of sensors, for which $\mathbf{w}_{\tau+1}$, minimizes $H_{\tau+1}(\boldsymbol{f}|\check{\boldsymbol{s}}_{\tau+1}, \boldsymbol{z}, \boldsymbol{\theta})$. Given $\check{\boldsymbol{s}}_\tau$, we write

$$H_{\tau+1}(\boldsymbol{f}|\check{\boldsymbol{s}}_{\tau+1}, \boldsymbol{z}, \boldsymbol{\theta}) = H_\tau(\boldsymbol{f}|\check{\boldsymbol{s}}_\tau, \boldsymbol{z}, \boldsymbol{\theta})$$
$$- \sum_{\boldsymbol{z}' \in \mathcal{Z}} \int_{\boldsymbol{\theta}', \check{\boldsymbol{s}}'_{\tau+1}} p(\check{\boldsymbol{s}}'_{\tau+1}, \boldsymbol{z}', \boldsymbol{\theta}')q(\boldsymbol{z}', \boldsymbol{\theta}', \mathbf{w}_{\tau+1})d\boldsymbol{\theta}' d\check{\boldsymbol{s}}'_{\tau+1}$$
$$(50)$$

---

**Algorithm 5** Adaptive Bayesian radio tomography

**Input:** $\boldsymbol{z}^{(0)}, \boldsymbol{\theta}^{(0)}, \check{\boldsymbol{s}}^{(0)}, N_{\text{CL}}, N_{\text{Burn-in}},$ and $N_{\text{Iter}}$.
1: Set $\check{\boldsymbol{s}}_0 = \check{\boldsymbol{s}}^{(0)}$
2: **for** $\tau = 0, 1, \ldots$ **do**
3:    Obtain $\mathcal{S}^{(\tau)}$ via Alg. 1($\boldsymbol{z}^{(0)}, \boldsymbol{\theta}^{(0)}, \check{\boldsymbol{s}}_\tau, N_{\text{CL}}, N_{\text{Burn-in}}, N_{\text{Iter}}$)
4:    Obtain $\hat{\boldsymbol{z}}^{(\tau)}_{\text{MAP}}$ from (39) by using $\mathcal{S}^{(\tau)}$
5:    Obtain $\hat{\boldsymbol{f}}^{(\tau)}_{\text{MMSE}}$ from (40) by using $\hat{\boldsymbol{z}}^{(\tau)}_{\text{MAP}}$ and $\mathcal{S}^{(\tau)}$
6:    Obtain $\hat{\boldsymbol{\theta}}^{(\tau)}_{\text{MMSE}}$ from (41)–(44) by using $\mathcal{S}^{(\tau)}$
7:    Calculate $\bar{u}(\mathbf{w})$ in (52) for $\mathbf{w} \in \mathcal{W}_{\tau+1}$ by using $\mathcal{S}^{(\tau)}$
8:    Collect $\check{s}_{\tau+1}$ from sensors associated with $\max \bar{u}(\mathbf{w})$
9:    Construct $\check{\boldsymbol{s}}_{\tau+1} = [\check{\boldsymbol{s}}^\top_\tau, \check{s}_{\tau+1}]^\top$
10:   Set $\boldsymbol{z}^{(0)} = \hat{\boldsymbol{z}}^{(\tau)}_{\text{MAP}}$ and $\boldsymbol{\theta}^{(0)} = \hat{\boldsymbol{\theta}}^{(\tau)}_{\text{MMSE}}$
11: **end for**

---

with $q(\boldsymbol{z}, \boldsymbol{\theta}, \mathbf{w}) := \ln\left(1 + (\sigma_\nu^2)^{-1}\mathbf{w}^\top \boldsymbol{\Sigma}_{f|z}\mathbf{w}\right)/2$, and seek $\mathbf{w}_{\tau+1}$ by solving

$$(\text{P1}) \quad \max_{\mathbf{w} \in \mathcal{W}_{\tau+1}} \mathbb{E}_{\boldsymbol{z},\boldsymbol{\theta}|\check{\boldsymbol{s}}_\tau}[q(\boldsymbol{z}, \boldsymbol{\theta}, \mathbf{w})]$$
$$= \sum_{\boldsymbol{z}' \in \mathcal{Z}} \int_{\boldsymbol{\theta}'} p(\boldsymbol{z}', \boldsymbol{\theta}'|\check{\boldsymbol{s}}_\tau)q(\boldsymbol{z}', \boldsymbol{\theta}', \mathbf{w})d\boldsymbol{\theta}' \quad (51)$$

where $\mathcal{W}_{\tau+1}$ is a set of weight vectors found from locations of available sensing radio pairs at time slot $\tau + 1$ (see Appendix B for derivation of (P1)). Note that solving (P1) to find $\mathbf{w}_{\tau+1}$ does not require $p(\boldsymbol{z}', \boldsymbol{\theta}'|\check{\boldsymbol{s}}_{\tau+1})$, which means the joint posterior in (15) does not need to be retrained for adaptive data acquisition.

Apparently, solving (P1) is not an easy task since evaluating $\mathbb{E}_{\boldsymbol{z},\boldsymbol{\theta}|\check{\boldsymbol{s}}_\tau}[q(\boldsymbol{z}, \boldsymbol{\theta}, \mathbf{w})]$ is intractable especially for large $N_g$ since $|\mathcal{Z}| = 2^{N_g}$. Fortunately, the samples from Alg. 1 can be used to approximate

$$\mathbb{E}_{\boldsymbol{z},\boldsymbol{\theta}|\check{\boldsymbol{s}}_\tau}[q(\boldsymbol{z}, \boldsymbol{\theta}, \mathbf{w})] \simeq \frac{1}{|\mathcal{S}^{(\tau)}|} \sum_{l=N_{\text{Burn-in}}+1}^{N_{\text{Iter}}} q(\boldsymbol{z}^{(l)}, \boldsymbol{\theta}^{(l)}, \mathbf{w}) =: \bar{u}(\mathbf{w}).$$
$$(52)$$

Therefore, $\check{s}_{\tau+1}$ can be obtained from the pair of sensors corresponding to $\mathbf{w}$ with the maximum value of $\bar{u}(\mathbf{w})$ in (52).

The steps involved for adaptive Bayesian radio tomography are listed in Alg. 5.

**Remark 2 (Mini-batch setup).** The proposed adaptive data acquisition method can be easily extended to a mini-batch setup of size $N_{\text{Batch}}$ per time slot $\tau$ as follows: i) find weight vectors $\{\mathbf{w}^{(i)}\}_{i=1}^{N_{\text{Batch}}} \subset \mathcal{W}_{\tau+1}$ associated with $N_{\text{Batch}}$ largest values of $\bar{u}(\mathbf{w})$ in (52), and collect the corresponding measurements $\{\check{s}^{(i)}_{\tau+1}\}_{i=1}^{N_{\text{Batch}}}$ (steps 7–8 in Alg. 5); and ii) aggregate those measurements below $\check{\boldsymbol{s}}_\tau$ to construct $\check{\boldsymbol{s}}_{\tau+1} := [\check{\boldsymbol{s}}^\top_\tau, \check{s}^{(1)}_{\tau+1}, \ldots, \check{s}^{(N_{\text{Batch}})}_{\tau+1}]^\top$ (step 9 in Alg. 5). Numerical tests will be performed to assess the mini-batch operation of Alg. 5.

## IV. NUMERICAL TESTS

Performance of the proposed algorithms was assessed through numerical tests using both synthetic and real datasets. A few existing methods were also tested for comparison, including the ridge-regularized SLF estimate given by $\hat{\boldsymbol{f}}_{\text{LS}} = (\mathbf{W}\mathbf{W}^\top + \mu_f \mathbf{C}_f^{-1})^{-1}\mathbf{W}\check{\boldsymbol{s}}_t$ [9], where $\mathbf{C}_f$ is a spatial covariance matrix modeling the similarity between points $\tilde{\mathbf{x}}_i$ and $\tilde{\mathbf{x}}_j$ in area $\mathcal{A}$. We further tested the total variation (TV)-regularized LS scheme in [26], which solves the regularized problem in (6) with

$$\mathcal{R}(\boldsymbol{f}) = \sum_{i=1}^{N_x-1} \sum_{j=1}^{N_y} |f_{i+1,j} - f_{i,j}| + \sum_{i=1}^{N_x} \sum_{j=1}^{N_y-1} |f_{i,j+1} - f_{i,j}|, \quad (53)$$

where $\mathbf{F} := \mathrm{unvec}(\boldsymbol{f}) \in \mathbb{R}^{N_x \times N_y}$, and $f_{i,j}$ denotes the $(i,j)$-th element of $\mathbf{F}$. As a competing alternative of the proposed adaptive sampling, simple *random sampling* was considered for both regularized LS estimators, by selecting $\{\check{s}_{\tau+1}^{(i)}\}_{i=1}^{N_{\mathrm{Batch}}}$ $\forall \tau$ uniformly at random. Particularly, Alg. 5 after replacing steps 7–8 with random sampling is named as the *non-adaptive Bayesian algorithm*, and will be compared with the proposed method throughout synthetic and real data tests.

### A. Test with synthetic data

This section validates the proposed algorithm through synthetic tests. Random tomographic measurements were taken by $N = 120$ sensors uniformly deployed on boundaries of $\mathcal{A} := [0.5, 40.5] \times [0.5, 40.5]$, from which the SLF defined over a grid $\{\tilde{\mathbf{x}}_i\}_{i=1}^{1,600} := \{1, \ldots, 40\}^2$ was reconstructed. To generate the ground-truth SLF $\boldsymbol{f}_0$, the hidden label field $\boldsymbol{z}_0$ was obtained first via the Metropolis algorithm [21] by using the prior of $\boldsymbol{z}$ in (8) with $\beta = 1.3$. Afterwards, $\boldsymbol{f}_0$ was constructed to have $f(\tilde{\mathbf{x}}_i) \sim \mathcal{N}(0.2, 1)$ $\forall \tilde{\mathbf{x}}_i \in \mathcal{A}_0$ and $f(\tilde{\mathbf{x}}_j) \sim \mathcal{N}(5, 0.2)$ $\forall \tilde{\mathbf{x}}_j \in \mathcal{A}_1$ resulting in $\boldsymbol{\theta}_f = [0.2, 5, 1, 0.2]^\top$, respectively, based on labels in $\boldsymbol{z}_0$. True $\mathbf{F}_0 := \mathrm{unvec}(\boldsymbol{f}_0) \in \mathbb{R}^{40 \times 40}$ and $\mathbf{Z}_0 := \mathrm{unvec}(\boldsymbol{z}_0) \in \{0,1\}^{40 \times 40}$ are depicted in Fig. 4 with sensor locations marked with crosses. The effects of calibration are not accounted for this section, meaning that $g_0$ and $\gamma$ are assumed to be known, and the fusion center directly uses shadowing measurements $\check{s}_\tau$. Under the mini-batch operation, each measurement $\check{s}_\tau^{(i)}$ $\forall \tau, i$ was generated according to (5), where $s_\tau$ was obtained by (4) with $w$ set to the normalized ellipse model in (3), while $\nu_\tau$ was set to follow zero-mean Gaussian with $\sigma_\nu^2 = 5 \times 10^{-2}$. To construct $\mathcal{W}_{\tau+1}$ per time slot $\tau$, $|\mathcal{W}_{\tau+1}| = 100$ pairs of sensors were uniformly selected at random with replacement. Then, $N_{\mathrm{Batch}} = 40$ shadowing measurements corresponding to $\{\mathbf{w}^{(i)}\}_{i=1}^{N_{\mathrm{Batch}}} \subset \mathcal{W}_{\tau+1}$ were collected to execute Alg. 5 for $\tau = 0, \ldots, 15$.

In all synthetic tests, the following simulation parameters were used: $N_{\mathrm{CL}} = 2$, $N_{\mathrm{Burn-in}} = 200$, $N_{\mathrm{Iter}} = 500$, and $\sigma_q^2 = 0.03$ were used to run the proposed algorithm; and hyper-hyperparameters of $\boldsymbol{\theta}$ were set as listed in Table I. For initialization, $\boldsymbol{\theta}^{(0)}$ was set to have $\beta^{(0)} = 0.1$, $\boldsymbol{\mu}_{f_k}^{(0)} = [m_0, m_1]^\top$, and randomly initialized $\sigma_\nu^2$ and $\boldsymbol{\sigma}_{f_k}^2$. Vector $\boldsymbol{z}^{(0)}$ was obtained by drawing $z_i^{(0)} \sim \mathrm{Bern}(0.5)$ for $i = 1, \ldots, N_g$, where $\mathrm{Bern}(0.5)$ denotes the Bernoulli distribution with mean equal to 0.5. Furthermore, $\check{s}^{(0)}$ was collected from randomly selected 100 pairs of sensors. For both regularized LS estimators, $\mu_f$ was found to minimize $\mathbb{E}[\|\boldsymbol{f}_0 - \hat{\boldsymbol{f}}\|_2^2]$ through grid-search.

The first experiment was performed to validate the efficacy of Alg. 5. The estimates $\hat{\mathbf{F}} = \mathrm{unvec}(\hat{\boldsymbol{f}})$ and $\hat{\mathbf{Z}} = \mathrm{unvec}(\hat{\boldsymbol{z}})$ at $\tau = 15$ are displayed in Figs. 5c and 5d, respectively, together with the estimated SLFs from the regularized-LS estimators in Figs. 5a and 5b. The most satisfactory result was obtained by the proposed method since piecewise homogeneous regions of the SLF were separately reconstructed by introducing the hidden label field.

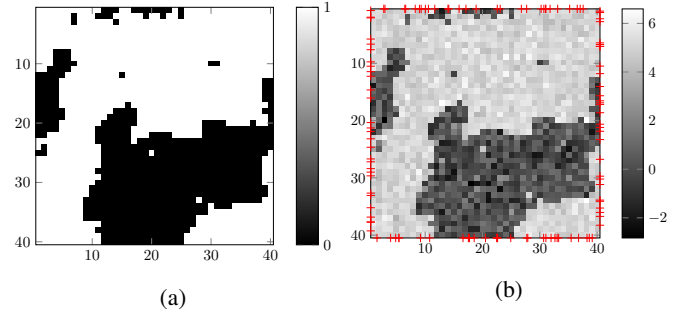

Fig. 4: True fields for synthetic tests: (a) hidden label field $\mathbf{Z}_0$ and (b) spatial loss field $\mathbf{F}_0$ with $N = 120$ sensor locations marked with crosses.

TABLE I: Hyper-hyperparameters of $\boldsymbol{\theta}$ for synthetic tests.

| $\beta_{\max}$ | $m_0$ | $m_1$ | $\sigma_k^2, \forall k$ | $a_\nu$ | $b_\nu$ | $a_k, \forall k$ | $b_k, \forall k$ |
|---|---|---|---|---|---|---|---|
| 10 | 0.5 | 4.5 | 0.1 | 0.1 | 0.1 | 0.1 | 0.1 |

TABLE II: True $\boldsymbol{\theta}$ and estimated $\hat{\boldsymbol{\theta}}$ via Alg. 5 (setting of Figs. 5c and 5d); and non-adaptive Bayesian algorithm (setting of Figs. 5e and 5f ) averaged over 20 independent Monte Carlo runs.

| $\boldsymbol{\theta}$ | True | Est. (Alg. 5) | Est. (non-adaptive) |
|---|---|---|---|
| $\beta$ | 1.3 | $1.309 \pm 2 \times 10^{-2}$ | $1.309 \pm 3 \times 10^{-2}$ |
| $\sigma_\nu^2$ | 0.05 | $0.058 \pm 10^{-2}$ | $0.053 \pm 1.3 \times 10^{-2}$ |
| $\mu_{f_0}$ | 0.2 | $0.289 \pm 2 \times 10^{-2}$ | $0.289 \pm 1.8 \times 10^{-2}$ |
| $\mu_{f_1}$ | 5 | $4.996 \pm 7 \times 10^{-3}$ | $4.996 \pm 7 \times 10^{-3}$ |
| $\sigma_{f_0}^2$ | 1 | $0.931 \pm 5 \times 10^{-2}$ | $0.94 \pm 9.8 \times 10^{-2}$ |
| $\sigma_{f_1}^2$ | 0.2 | $0.198 \pm 2 \times 10^{-2}$ | $0.193 \pm 2.8 \times 10^{-2}$ |

To test the proposed adaptive data acquisition method, $\hat{\mathbf{F}}$ and $\hat{\mathbf{Z}}$ reconstructed by the non-adaptive Bayesian algorithm are shown in Figs. 5e and 5f, respectively. Comparison between Figs. 5c and 5e visually demonstrates that improved SLF reconstruction performance could be achieved through adaptive data acquisition with the same number of measurements. Accuracy of $\hat{\boldsymbol{z}}$ was also quantitatively measured by the labeling-error, defined as $\|\boldsymbol{z}_0 - \hat{\boldsymbol{z}}\|_1/N_g$. Fig. 6 displays the progression of the labeling-error averaged over 20 independent Monte Carlo runs. It shows that the proposed adaptive method consistently outperforms the non-adaptive one, which implies that selection of informative measurements to decrease uncertainty of $\boldsymbol{f}$ given current estimates of $\boldsymbol{z}$ and $\boldsymbol{\theta}$ could lead to more accurate estimates of $\boldsymbol{f}$ and $\boldsymbol{z}$ in the next time slot.

Meanwhile, average estimates of $\boldsymbol{\theta}$ and associated standard deviation denoted with $\pm$ are listed in Table II, where every hyperparameter was accurately estimated. Together with the result in Fig. 5, the accurate estimates of the hyperparameters confirm that the proposed method can faithfully capture patterns of objects in area of interest, and also reveal the underlying statistical properties.

The rest of this section tests the performance of the proposed algorithm in channel-gain cartography. To this end, the same setting used to produce Figs. 5c and 5d was adopted. From the estimate $\hat{\boldsymbol{f}}_{\text{MMSE}}$ obtained through Alg. 5, an estimate of the shadowing attenuation $\hat{s}(\mathbf{x}, \mathbf{x}')$ between two arbitrary points $\mathbf{x}$ and $\mathbf{x}'$ in $\mathcal{A}$ is obtained through (4) by replacing $\boldsymbol{f}$ with $\hat{\boldsymbol{f}}_{\text{MMSE}}$. Subsequently, an estimate of the channel-gain $\hat{g}(\mathbf{x}, \mathbf{x}')$ is obtained after substituting $\hat{s}(\mathbf{x}, \mathbf{x}')$ into (1).

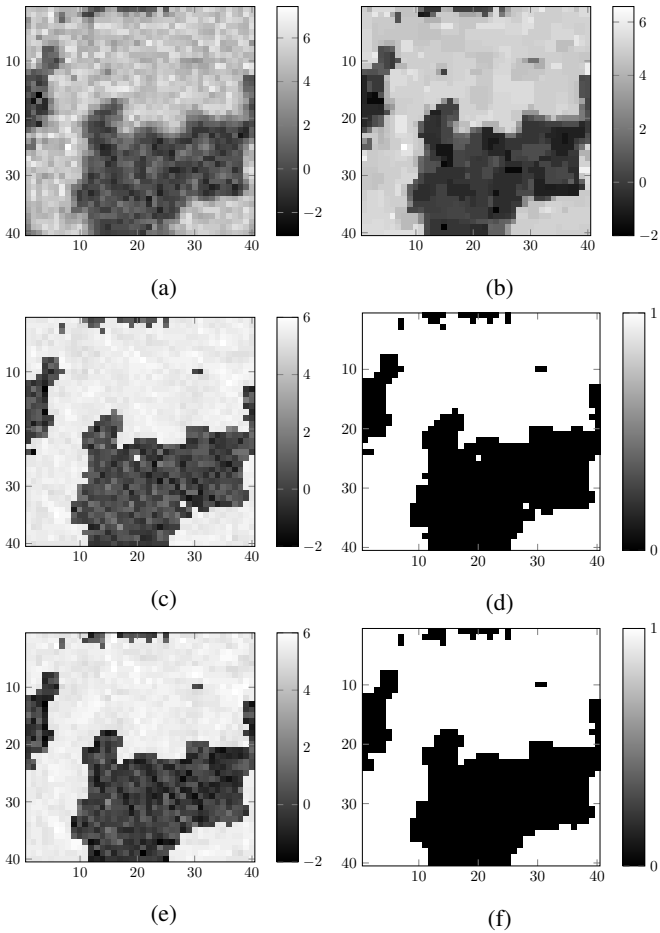

Fig. 5: Estimated SLFs $\hat{\mathbf{F}}$ at $\tau = 15$ (with 700 measurements) via (a) ridge-regularized LS ($\mu_f = 10^{-6}$ and $\mathbf{C}_f = \mathbf{I}_{1,600}$); (b) TV-regularized LS ($\mu_f = 10^{-4}$); (c) Alg. 5 through (d) estimated hidden label field $\hat{\mathbf{Z}}$; and (e) non-adaptive Bayesian algorithm, through (f) estimated $\hat{\mathbf{Z}}$.

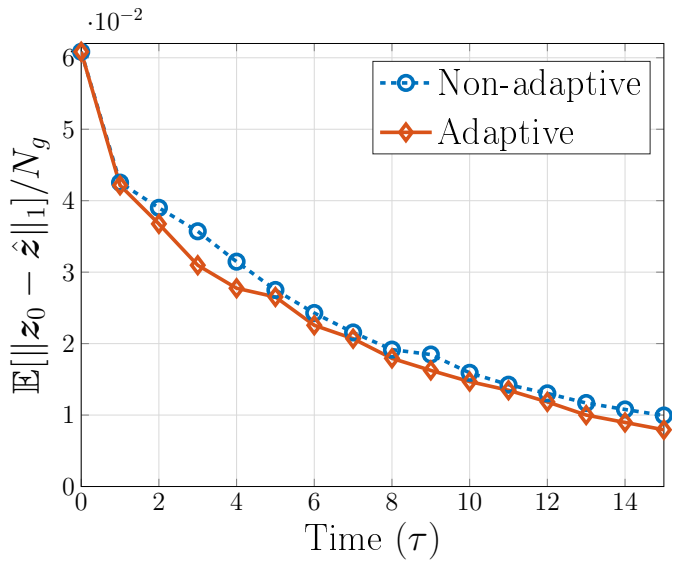

Fig. 6: Progression of error in estimation of $\mathbf{z}$.

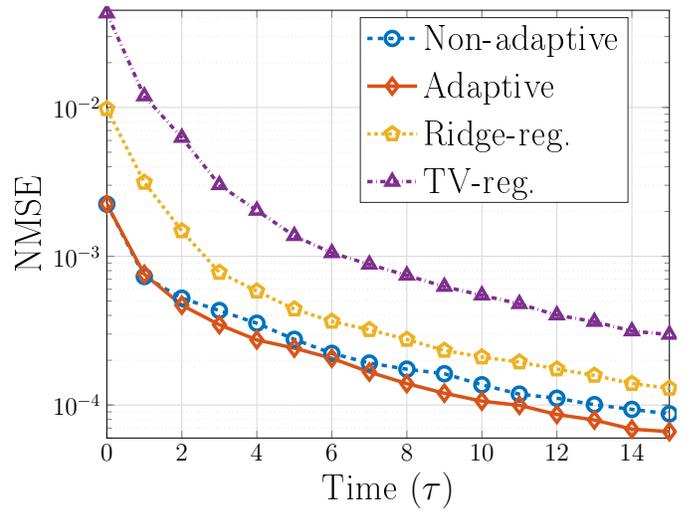

Fig. 7: Progression of channel-gain estimation error.

TABLE III: Hyper-hyperparameters of $\boldsymbol{\theta}$ for real data tests.

| $\beta_{\max}$ | $m_0$ | $m_1$ | $\sigma_k^2, \forall k$ | $a_\nu$ | $b_\nu$ | $a_k, \forall k$ | $b_k, \forall k$ |
|---|---|---|---|---|---|---|---|
| 2 | 0 | 1 | 0.01 | 1 | 0.01 | 0.01 | 0.01 |

Since $g_0$ and $\gamma$ are known, obtaining $s(\mathbf{x}, \mathbf{x}')$ amounts to finding $g(\mathbf{x}, \mathbf{x}')$; cf. (1). This suggests adopting a performance metric quantifying the mismatch between $s(\mathbf{x}, \mathbf{x}')$ and $\hat{s}(\mathbf{x}, \mathbf{x}')$, using the normalized mean-square error

$$\text{NMSE} := \frac{\mathbb{E}\big[\int_{\mathcal{A}} \big(s(\mathbf{x}, \mathbf{x}') - \hat{s}(\mathbf{x}, \mathbf{x}')\big)^2 d\mathbf{x} d\mathbf{x}'\big]}{\mathbb{E}\big[\int_{\mathcal{A}} s^2(\mathbf{x}, \mathbf{x}') d\mathbf{x} d\mathbf{x}'\big]}$$

where the expectation is over the set $\{\mathbf{x}_n\}_{n=1}^N$ of sensor locations and realizations of $\{\nu_\tau\}_\tau$. Simulations estimated the expectations by averaging over 20 independent Monte Carlo runs. The integrals are approximated by averaging the integrand over 300 pairs of $(\mathbf{x}, \mathbf{x}')$ chosen independently and uniformly at random over the boundary of $\mathcal{A}$.

Fig. 7 compares the NMSE of the proposed method with those of the competing alternatives using the settings in Fig. 5. Evidently, the proposed method achieves the lowest NMSE for every $\tau$. Observe that both Bayesian approaches outperform the regularized LS methods, which suggests the proposed method as a viable alternative of a conventional solution adopted for radio tomography and channel-gain cartography.

### B. Test with real data

This section validates the proposed method using the real data set in [9]. The test setup is depicted in Fig. 8, where $\mathcal{A} = [0.5, 20.5] \times [0.5, 20.5]$ is a square with sides of 20 feet (ft), over which a grid $\{\tilde{\mathbf{x}}_i\}_{i=1}^{1,681} := \{1, \ldots, 41\}^2$ of $N_g = 1,681$ points is defined. A collection of $N = 80$ sensors measure the channel attenuation at 2.425 GHz between pairs of sensor positions, marked with the $N = 80$ crosses in Fig. 8. To estimate $g_0$ and $\gamma$ using the approach in [9], a first set of $2,400$ measurements was obtained before placing the artificial structure in Fig. 8. Estimates $\hat{g}_0 = 54.6$ (dB) and $\hat{\gamma} = 0.276$ were obtained during the calibration step. Afterwards, the structure comprising one pillar and six walls of different materials was assembled, and $T = 2,380$ measurements $\{\check{g}_{\tau'}\}_{\tau'=1}^{T}$ were acquired. Then, the calibrated measurements $\{\check{s}_{\tau'}\}_{\tau'=1}^{T}$ were obtained from $\{\check{g}_{\tau'}\}_{\tau'=1}^{T}$ by substituting $\hat{g}_0$ and $\hat{\gamma}$ into (5). In addition, $\{\mathbf{w}_{\tau'}\}_{\tau'=1}^{T}$ were constructed with $w$ in (3) by using known locations of sensor pairs. Note that $\tau'$ is introduced to distinguish indices of the real data from $\tau$ used to index time slots in numerical tests.

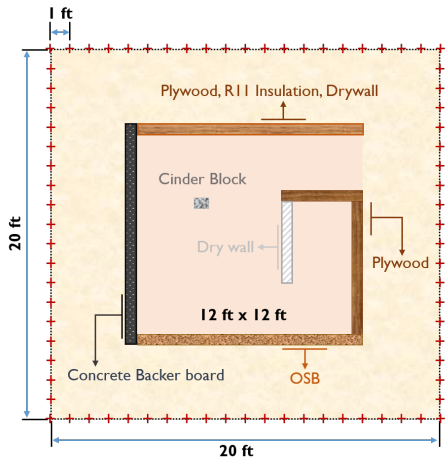

Fig. 8: Configuration of the testbed with $N = 80$ sensor locations marked with crosses.

We randomly selected $1,380$ measurements from $\{\check{s}_{\tau'}\}_{\tau'=1}^{T}$ to initialize $\check{\mathbf{s}}^{(0)}$, and used the remaining $1,000$ measurements to run the proposed algorithm under the mini-batch operation for $\tau = 0, \ldots, 5$, where every $\mathcal{W}_{\tau+1}$ was formed by uniformly selecting $|\mathcal{W}_{\tau+1}| = 200$ weight vectors at random from $\{\mathbf{w}_{\tau'}\}_{\tau'}$ associated with the remaining $1,000$ measurements without replacement. Parameters of the proposed algorithm were set to, $N_{\text{CL}} = 2$, $N_{\text{Burn-in}} = 300$, $N_{\text{Iter}} = 1,000$, $\sigma_q^2 = 10^{-5}$, and the hyper-hyperparameters of $\boldsymbol{\theta}$ used are listed in Table III. For initialization, $\mathbf{z}^{(0)}$ was found by drawing $z_i^{(0)} \sim \text{Bern}(0.5)\ \forall i$. Vector $\boldsymbol{\theta}^{(0)}$ was set to have $\beta^{(0)} = 0.1$ and $\boldsymbol{\mu}_{f_k}^{(0)} = [m_0, m_1]^\top$, while $\sigma_\nu^2$ and $\boldsymbol{\sigma}_{f_k}^2$ were initialized at random.

Following [1], [9], a spatial covariance matrix was used for $\mathbf{C}_f$ of the ridge-regularized LS estimator, which models the similarity between points $\tilde{\mathbf{x}}_i$ and $\tilde{\mathbf{x}}_j$ as $[\mathbf{C}_f]_{ij} = \sigma_s^2 \exp[-\|\tilde{\mathbf{x}}_i - \tilde{\mathbf{x}}_j\|_2 / \kappa]$ [1], with $\sigma_s^2 = \kappa = 1$, and $\mu_f = 6 \times 10^{-2}$; see also [27]. On the other hand, the TV-regularized LS estimator was tested with $\mu_f = 4.3$ used in [27].

Fig. 9 displays estimated SLFs $\hat{\mathbf{F}}$ and associated hidden fields $\hat{\mathbf{Z}}$ at $\tau = 5$ obtained by the proposed method and its competing alternatives. The pattern of the artificial structure is clearly delineated on $\hat{\mathbf{F}}$ in Fig. 9c estimated by the proposed method, while the regularized LS estimators are not able to capture such pattern without post-processing of the estimated SLFs in Figs. 9a and 9b. Although the non-adaptive Bayesian algorithm reconstructed the visually satisfying SLF for radio tomography as shown in Fig. 9e, $\hat{\mathbf{F}}$ from the proposed method depicts the artificial structure more clearly; see e.g., object patterns in Figs. 9c and 9e corresponding to the dry wall in Fig. 8. As a benchmark, an one-shot estimate of the SLF, denoted as $\hat{\mathbf{F}}_{\text{full}}$, is also displayed in Fig. 9g, which was obtained via Alg. 5 by using the entire set of $2,380$ measurements. Comparison of $\hat{\mathbf{F}}$ in Fig. 9c with $\hat{\mathbf{F}}_{\text{full}}$ shows that the proposed algorithm enables one to reconstruct the SLF close to the benchmark by using fewer, but more informative measurements.

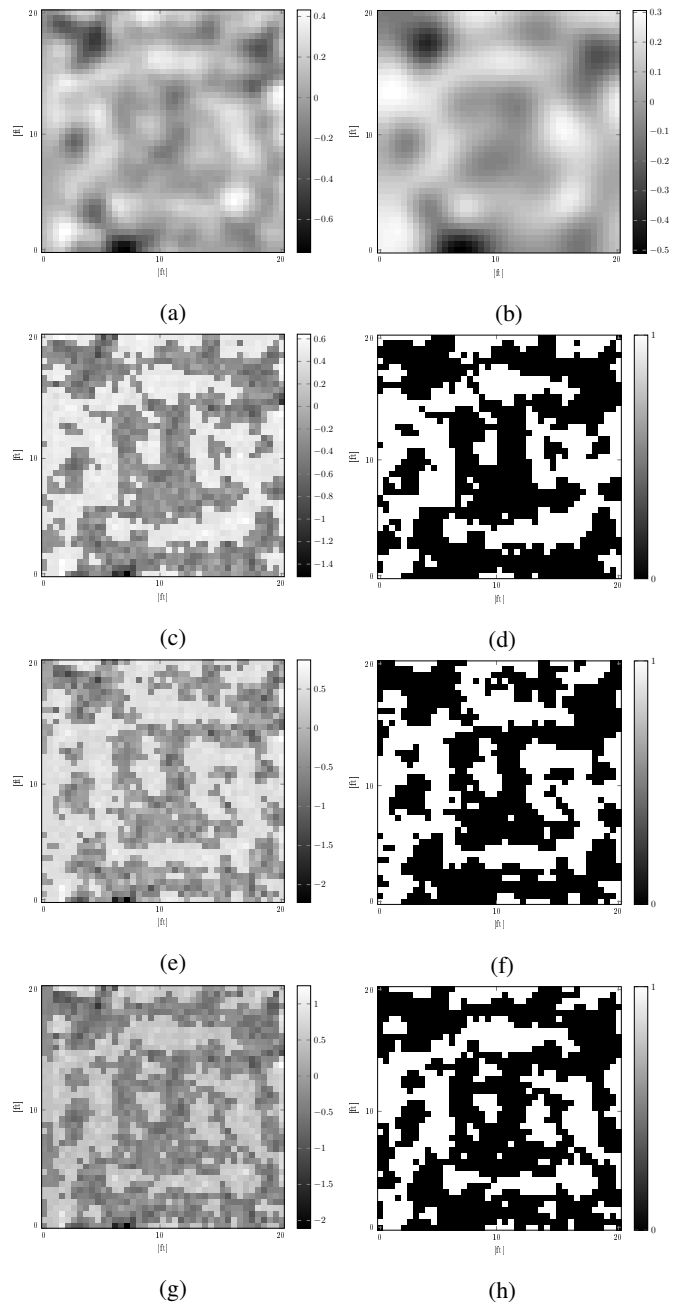

Fig. 9: Estimated SLFs $\hat{\mathbf{F}}$ at $\tau = 5$ (with $1,880$ measurements) via (a) ridge-regularized LS; (b) TV-regularized LS; (c) Alg. 5 through (d) estimated hidden label field $\hat{\mathbf{Z}}$; and (e) non-adaptive Bayesian algorithm, through (f) estimated $\hat{\mathbf{Z}}$, together with one-shot estimates (g) $\hat{\mathbf{F}}_{\text{full}}$ and (h) $\hat{\mathbf{Z}}_{\text{full}}$ obtained by using the full dataset (with $2,380$ measurements) via Alg. 5.

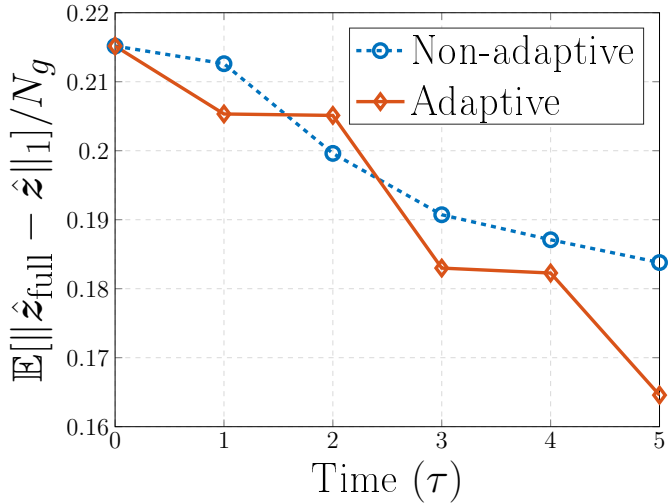

Fig. 10: Progression of a mismatch between $\hat{z}$ and $\hat{z}_{\text{full}}$.

TABLE IV: Estimated $\hat{\boldsymbol{\theta}}$ via benchmark algorithm (setting of Figs. 9g and 9h); Alg. 5 (setting of Figs. 9c and 9d); and non-adaptive Bayesian algorithm (setting of Figs. 9e and 9f), averaged over 20 independent Monte Carlo runs.

| $\boldsymbol{\theta}$ | Est. (benchmark) | Est. (Alg. 5) | Est. (non-adaptive) |
|---|---|---|---|
| $\beta$ | $0.499 \pm 2 \times 10^{-4}$ | $0.5 \pm 5 \times 10^{-4}$ | $0.5 \pm 6 \times 10^{-4}$ |
| $\sigma_\nu^2$ | $9.984 \pm 0.05$ | $10.60 \pm 0.20$ | $9.957 \pm 0.23$ |
| $\mu_{f_0}$ | $-0.275 \pm 0.02$ | $-0.278 \pm 0.02$ | $-0.301 \pm 0.03$ |
| $\mu_{f_1}$ | $0.463 \pm 0.03$ | $0.447 \pm 0.03$ | $0.504 \pm 0.03$ |
| $\sigma_{f_0}^2$ | $0.629 \pm 0.12$ | $0.457 \pm 0.13$ | $0.456 \pm 0.22$ |
| $\sigma_{f_1}^2$ | $0.171 \pm 0.10$ | $0.145 \pm 0.10$ | $0.325 \pm 0.43$ |

The second experiment investigated the efficacy of the proposed adaptive data acquisition method in estimating $z$. By considering $\hat{\mathbf{Z}}_{\text{full}} = \text{unvec}(\hat{z}_{\text{full}})$ in Fig. 9h as a benchmark, the labeling error $\|\hat{z}_{\text{full}} - \hat{z}\|_1 / N_g$ was used as a performance metric. Fig. 10 compares the labeling error of the proposed method with that of the non-adaptive algorithm, which are averaged over 20 independent Monte Carlo runs. The proposed method exhibits lower labeling errors than the non-adaptive one except when $\tau = 2$. This illustrates that the proposed data acquisition criterion delineates object patterns more accurately while also reducing the measurement collection cost.

To corroborate the hyperparameter estimation capability of the proposed algorithm, the estimates of $\boldsymbol{\theta}$ averaged over 20 independent Monte Carlo runs were listed in Table IV. The estimate $\hat{\boldsymbol{\theta}}$ obtained by using the full dataset was considered as a benchmark, to demonstrate that the proposed method estimates $\boldsymbol{\theta}$ closer to the benchmark.

The last simulation assesses the performance of the proposed algorithm and competing alternatives for channel-gain cartography. The same set of shadowing measurements and simulation setup as in first simulations of this section were used. A channel-gain map is constructed to portray the gain between any point in the map, and a fixed receiver location $\mathbf{x}_{\text{rx}}$. Particularly, the proposed algorithm is executed and estimates $\{\hat{s}(\tilde{\mathbf{x}}_i, \mathbf{x}_{\text{rx}})\}_{i=1}^{N_g}$ are obtained by substituting $\hat{f}$ and $w$ into (4). Subsequently, $\{\hat{g}(\tilde{\mathbf{x}}_i, \mathbf{x}_{\text{rx}})\}_{i=1}^{N_g}$ are obtained by substituting $\{\hat{s}(\tilde{\mathbf{x}}_i, \mathbf{x}_{\text{rx}})\}_{i=1}^{N_g}$ into (1) with $\hat{g}_0$ and $\hat{\gamma}$. After defining $\hat{\mathbf{g}} := [\hat{g}(\tilde{\mathbf{x}}_1, \mathbf{x}_{\text{rx}}), \ldots, \hat{g}(\tilde{\mathbf{x}}_{N_g}, \mathbf{x}_{\text{rx}})]^\top$, one can construct the channel-gain map $\hat{\mathbf{G}} := \text{unvec}(\hat{\boldsymbol{g}})$ with the receiver located at $\mathbf{x}_{\text{rx}}$.

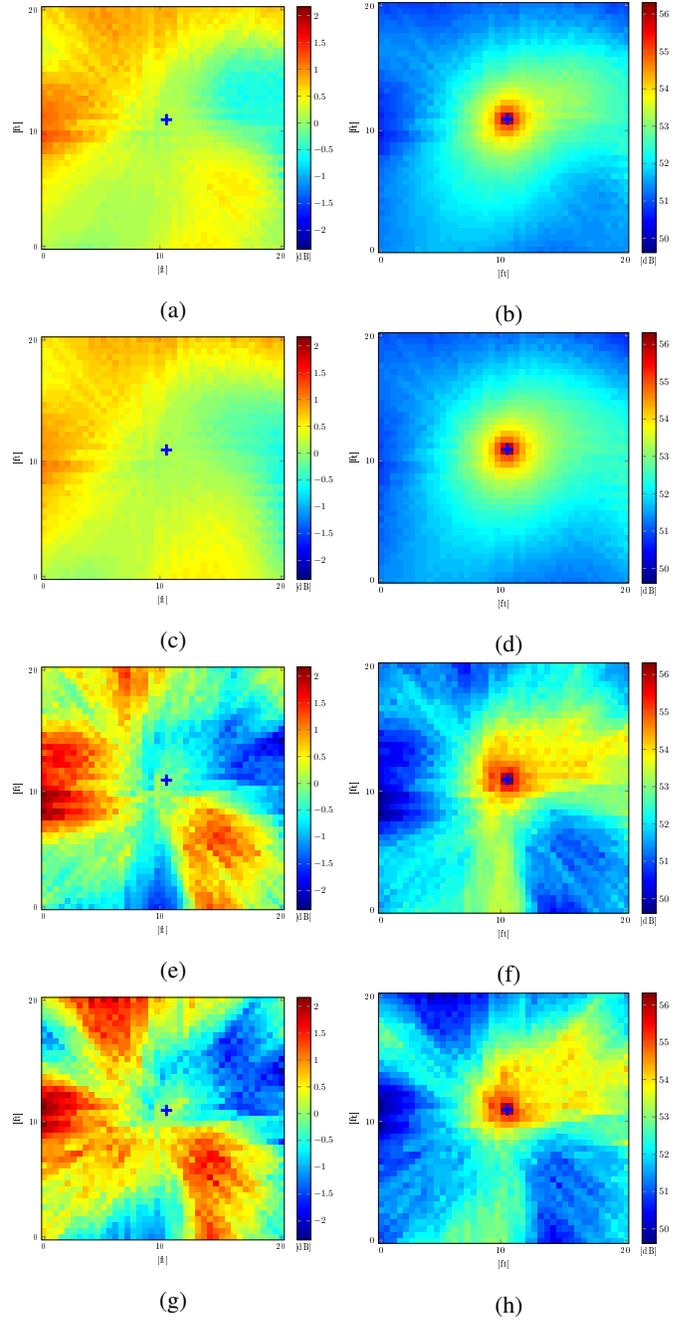

Fig. 11: Estimated shadowing maps $\hat{\mathbf{S}}$ and corresponding channel-gain maps $\hat{\mathbf{G}}$ at $\tau = 5$ via (a)-(b) ridge-regularized LS (setting of Fig. 9a); (c)-(d) TV-regularized LS (setting of Fig. 9b); (e)-(f) Alg. 5 (setting of Fig. 9c); and (g)-(h) non-adaptive Bayesian algorithm (setting of Fig. 9e), with the receiver location at $\mathbf{x}_{\text{rx}} = (10.3, 10.7)$ (ft) marked with the blue cross.

Let $\hat{\mathbf{S}} := \text{unvec}(\hat{\boldsymbol{s}})$ denote the shadowing map with $\hat{\boldsymbol{s}} := [\hat{s}(\tilde{\mathbf{x}}_1, \mathbf{x}_{\text{rx}}), \ldots, \hat{s}(\tilde{\mathbf{x}}_{N_g}, \mathbf{x}_{\text{rx}})]^\top$. Fig. 11 displays the estimated shadowing maps $\hat{\mathbf{S}}$ and corresponding channel-gain maps $\hat{\mathbf{G}}$, obtained via various methods, when the receiver is located at $\mathbf{x}_{\text{rx}} = (10.3, 10.7)$ (ft) marked by the cross. In all channel-gain maps in Fig. 11, stronger attenuation is observed when a signal passes through either more building materials (bottom-right side of $\hat{\mathbf{G}}$), or the concrete wall (left side of $\hat{\mathbf{G}}$). In contrast, only the channel-gain maps in Figs. 11f and 11h reconstructed by the Bayesian methods exhibit less attenuation along the entrance of the artificial objects (top-right side of $\hat{\mathbf{G}}$), while channel-gain tends to drop quickly within the vicinity of the receiver in the channel-gain maps obtained by the regularized LS estimators, as shown in Figs. 11b and 11d. This stems from the fact that free space and objects are more distinctively delineated in $\hat{\mathbf{F}}$ by the Bayesian approaches. Note that slightly different observations were made in Figs. 11f and 11h since the shadowing map in Fig. 11g introduces stronger attenuation in free space below the receiver, which would disagree with intuition. All in all, the simulation results confirm that our approach could provide more specific CSI of the propagation medium, and thus endow the operation of cognitive radio networks with more accurate interference management.

## V. Conclusions

This paper developed a novel adaptive Bayesian radio tomographic algorithm that estimates the spatial loss field of the radio tomographic model, which is of interest in imaging and channel-gain cartography applications, by using measurements adaptively collected based on the uncertainty sampling criterion. Different from conventional approaches, leveraging a hidden label field contributed to effectively account for inhomogeneities of the spatial loss field. The effectiveness of the novel algorithm was corroborated through extensive synthetic and real data experiments. Future research will include an online approach to Bayesian radio tomography to further reduce computational complexity.

## Appendix

### A. Distribution of the proportionality of $p(\boldsymbol{f}|\check{\mathbf{s}}_t, \boldsymbol{z}, \boldsymbol{\theta})$

Recalling that $p(\check{\mathbf{s}}_t|\boldsymbol{f}, \sigma_\nu^2) \sim \mathcal{N}(\mathbf{W}^\top \boldsymbol{f}, \sigma_\nu^2 \mathbf{I}_t)$ and $p(\boldsymbol{f}|\boldsymbol{z}, \boldsymbol{\theta}_f) \sim \mathcal{N}(\boldsymbol{\mu}_{f|z}, \boldsymbol{\Delta}_{f|z})$, one can expand $p(\boldsymbol{f}|\check{\mathbf{s}}_t, \boldsymbol{z}, \boldsymbol{\theta})$ in (22) to arrive at (cf. (23))

$$p(\boldsymbol{f}|\check{\mathbf{s}}_t, \boldsymbol{z}, \boldsymbol{\theta}) \propto p(\check{\mathbf{s}}_t|\boldsymbol{f}, \sigma_\nu^2) p(\boldsymbol{f}|\boldsymbol{z}, \boldsymbol{\theta}_f)$$
$$\propto \exp\left[-\frac{1}{2\sigma_\nu^2}\|\check{\mathbf{s}}_t - \mathbf{W}^\top \boldsymbol{f}\|_2^2 - \frac{1}{2}\|\boldsymbol{f} - \boldsymbol{\mu}_{f|z}\|_{\boldsymbol{\Delta}_{f|z}^{-1}}^2\right]$$
$$\propto \exp\left[-\frac{1}{2}\boldsymbol{f}^\top \boldsymbol{\Sigma}_{f|z,\boldsymbol{\theta},\check{\mathbf{s}}_t}^{-1} \boldsymbol{f} + \left(\frac{1}{\sigma_\nu^2}\check{\mathbf{s}}_t^\top \mathbf{W}^\top + \boldsymbol{\mu}_{f|z}^\top \boldsymbol{\Delta}_{f|z}^{-1}\right)\boldsymbol{f}\right]$$
$$= \exp\left[-\frac{1}{2}\boldsymbol{f}^\top \boldsymbol{\Sigma}_{f|z,\boldsymbol{\theta},\check{\mathbf{s}}_t}^{-1} \boldsymbol{f} + \check{\boldsymbol{\mu}}_{f|z,\boldsymbol{\theta},\check{\mathbf{s}}_t}^\top \boldsymbol{\Sigma}_{f|z,\boldsymbol{\theta},\check{\mathbf{s}}_t}^{-1} \boldsymbol{f}\right]$$
$$\propto \exp\left[-\frac{1}{2}\|\boldsymbol{f} - \check{\boldsymbol{\mu}}_{f|z,\boldsymbol{\theta},\check{\mathbf{s}}_t}\|_{\boldsymbol{\Sigma}_{f|z,\boldsymbol{\theta},\check{\mathbf{s}}_t}^{-1}}^2\right], \quad (54)$$

which shows that the proportionality of $p(\boldsymbol{f}|\check{\mathbf{s}}_t, \boldsymbol{z}, \boldsymbol{\theta})$ follows $\mathcal{N}(\check{\boldsymbol{\mu}}_{f|z,\boldsymbol{\theta},\check{\mathbf{s}}_t}, \boldsymbol{\Sigma}_{f|z,\boldsymbol{\theta},\check{\mathbf{s}}_t})$. ∎

### B. Derivation of (P1)

At time slot $\tau$, we seek $\mathbf{w}_{\tau+1}$ minimizing $H_{\tau+1}(\boldsymbol{f}|\check{\mathbf{s}}_{\tau+1}, \boldsymbol{z}, \boldsymbol{\theta})$ in (50), which amounts to solving

$$\max_{\mathbf{w} \in \mathcal{W}_{\tau+1}} \sum_{\boldsymbol{z}' \in \mathcal{Z}} \int_{\boldsymbol{\theta}', \check{\mathbf{s}}'_{\tau+1}} p(\check{\mathbf{s}}'_{\tau+1}, \boldsymbol{z}', \boldsymbol{\theta}') q(\boldsymbol{z}', \boldsymbol{\theta}', \mathbf{w}) d\boldsymbol{\theta}' d\check{\mathbf{s}}'_{\tau+1}. \quad (55)$$

Then, one can show that

$$\int p(\check{\mathbf{s}}'_{\tau+1}, \boldsymbol{z}', \boldsymbol{\theta}') d\check{\mathbf{s}}'_{\tau+1}$$
$$= \int_{\check{\mathbf{s}}'_{\tau+1}} \int_{\boldsymbol{f}'} p(\check{\mathbf{s}}'_{\tau+1}, \boldsymbol{f}', \boldsymbol{z}', \boldsymbol{\theta}') d\boldsymbol{f}' d\check{\mathbf{s}}'_{\tau+1}$$
$$\stackrel{(e1)}{=} \iint p(\check{\mathbf{s}}'_{\tau+1}|\boldsymbol{f}', \boldsymbol{z}', \boldsymbol{\theta}') p(\check{\mathbf{s}}'_\tau|\boldsymbol{f}', \boldsymbol{z}', \boldsymbol{\theta}') p(\boldsymbol{f}', \boldsymbol{z}', \boldsymbol{\theta}') d\boldsymbol{f}' d\check{\mathbf{s}}'_{\tau+1}$$
$$= \iint p(\boldsymbol{f}', \boldsymbol{z}', \boldsymbol{\theta}'|\check{\mathbf{s}}'_\tau) p(\check{\mathbf{s}}'_\tau) d\boldsymbol{f}' d\check{\mathbf{s}}'_\tau = \int p(\boldsymbol{z}', \boldsymbol{\theta}'|\check{\mathbf{s}}'_\tau) p(\check{\mathbf{s}}'_\tau) d\check{\mathbf{s}}'_\tau \quad (56)$$

where $(e1)$ holds due to independence between $\check{\mathbf{s}}'_{\tau+1}$ and $\check{\mathbf{s}}'_\tau$ after conditioning on $\{\boldsymbol{f}, \boldsymbol{z}, \boldsymbol{\theta}\}$. By substituting (56) into (55) and recalling that $\check{\mathbf{s}}_\tau$ is given at time slot $\tau$, finding $\mathbf{w}_{\tau+1}$ boils down to solving

$$\max_{\mathbf{w} \in \mathcal{W}_{\tau+1}} \sum_{\boldsymbol{z}' \in \mathcal{Z}} \int_{\boldsymbol{\theta}'} p(\boldsymbol{z}', \boldsymbol{\theta}'|\check{\mathbf{s}}_\tau) q(\boldsymbol{z}', \boldsymbol{\theta}', \mathbf{w}) d\boldsymbol{\theta}', \quad (57)$$

which is (P1). ∎